\pdfoutput=1
\documentclass[conference]{IEEEtran}

\usepackage{graphicx}
\usepackage{amsmath,amssymb,amsfonts,bm}
\usepackage[T1]{fontenc}
\usepackage{lmodern}
\usepackage[english]{babel}

\usepackage{amsthm} 

\newtheoremstyle{breakthm}
  {\topsep}               
  {\topsep}               
  {\itshape}              
  {}                      
  {\bfseries}             
  {}                      
  {\newline}              
  {\thmname{#1}\thmnumber{ #2}\thmnote{ (\normalfont#3)}}

\theoremstyle{breakthm}
\newtheorem{proposition}{Proposition}
\theoremstyle{plain}

\numberwithin{assumption}{section}
\numberwithin{lemma}{section}
\numberwithin{theorem}{section}

\usepackage{enumitem}
\usepackage{array}
\usepackage{booktabs}
\usepackage{tabularx}
\usepackage{multirow}
\usepackage{makecell}
\usepackage{siunitx}
\usepackage{pgfplots}
\pgfplotsset{compat=1.18}
\usepackage[caption=false,font=footnotesize]{subfig}
\usepackage{adjustbox}
\usepackage{dblfloatfix}
\usepackage{placeins}

\usepackage{algorithm}
\usepackage{algpseudocode}
\algrenewcommand\algorithmicrequire{\textbf{Input:}}
\algrenewcommand\algorithmicensure{\textbf{Output:}}
\algrenewcommand\algorithmicreturn{\textbf{return}}

\usepackage[dvipsnames]{xcolor}
\usepackage[colorlinks=true,linkcolor=blue,citecolor=blue,urlcolor=blue]{hyperref}
\usepackage[most]{tcolorbox}
\tcbset{
  colback=gray!3,
  colframe=gray!60,
  left=6pt, right=6pt, top=6pt, bottom=6pt,
  boxsep=4pt
}

\usepackage{setspace}
\AtBeginDocument{\setstretch{1.2}}

\renewcommand{\arraystretch}{1.15}
\newcolumntype{Y}{>{\raggedright\arraybackslash}X}



\title{Comparative Analysis of Data Augmentation for Clinical ECG Classification with STAR}

\author{%
  \IEEEauthorblockN{Nader Nemati\IEEEauthorrefmark{1}\IEEEauthorrefmark{2}}
  \IEEEauthorblockA{\IEEEauthorrefmark{1}\textit{University of Turku}, Turku, Finland}
  \IEEEauthorblockA{\IEEEauthorrefmark{2}\textit{IEEE Machine Learning Member}, Turku, Finland}\\[-5mm]
  ORCID: \href{https://orcid.org/0000-0003-2548-6740}{0000-0003-2548-6740}
}

\begin{document}
\maketitle




\begin{abstract}
\begin{abstract}
Clinical 12-lead ECG classification remains difficult because of diverse recording conditions, overlapping pathologies, and pronounced label imbalance hinder generalization, while unconstrained augmentations risk distorting diagnostically critical morphology. In this study, \textbf{Sinusoidal Time--Amplitude Resampling (STAR)} is introduced as a beat-wise augmentation that operates strictly between successive R-peaks to apply controlled time warping and amplitude scaling to each R--R segment, preserving the canonical P--QRS--T order and leaving the head and tail of the trace unchanged.\\[1mm]
STAR is designed for practical pipelines and offers: (i) \emph{morphology-faithful variability} that broadens training diversity without corrupting peaks or intervals; (ii) \emph{source-resilient training}, improving stability across devices, sites, and cohorts without dataset-specific tuning; (iii) \emph{model-agnostic integration} with common 1D SE--ResNet-style ECG encoders backbone; and (iv) \emph{better learning on rare classes via beat-level augmentation}, reducing overfitting by resampling informative beats instead of duplicating whole records. In contrast to global crops, large shifts, or additive noise, STAR avoids transformations that suppress or misalign clinical landmarks.\\[1mm]
A complete Python implementation and a transparent training workflow are released, aligned with a source-aware, stratified five-fold protocol over a multi-institutional 12-lead corpus, thereby facilitating inspection and reuse. Taken together, STAR provides a simple and controllable augmentation for clinical ECG classification where trustworthy morphology, operational simplicity, and cross-source durability are essential.
\end{abstract}

\end{abstract}


\vspace{10mm}

\section{\textbf{Introduction}}

Cardiovascular diseases (CVDs) are a general category of chronic illnesses that pose significant risks to human well-being\cite{b1}. Among the various diagnostic tools available, electrocardiograms (ECGs) captured using the 12-lead technique play a critical role in detecting cardiac abnormalities and providing valuable diagnostic information about the heart\cite{b3,b20}. These 12-lead ECGs consist of twelve signals, each representing a specific angle correlated with different anatomical regions of the heart. The signals exhibit various waveforms, such as the T wave, P wave, and QRS complex, whose statistical and morphological features are essential health indicators and reveal symptoms related to cardiac conditions\cite{b20}.

Expert interpretation of ECG signals from multiple leads enables the localization and characterization of cardiac abnormalities, with cardiologists demonstrating a high level of accuracy in their assessments\cite{b3}. However, the proficiency of human interpretation is constrained by factors such as the availability of trained cardiologists and the time required to synthesize information from the 12-lead signal. To overcome these limitations, traditional machine learning approaches have played a crucial role in identifying salient features of the ECG signal, leading to successful applications in certain domains\cite{b4}. Nonetheless, there is potential for further enhancing interpretation accuracy by utilizing deep neural network methods. Unfortunately, the lack of labeled training data has hindered the application of these techniques to 12-lead ECGs.

Addressing this challenge, a clinical trial investigating the use of artificial-intelligence–enhanced ECG (AI-ECG) diagnosis has been conducted at Mayo Clinic. This trial aims to detect cardiovascular diseases and has shown promising benefits of employing AI-ECG. Deep learning techniques have proven effective for ECG classification in the research community, but their application is limited in practical clinical processes due to challenges arising from both deep learning techniques and ECG data. Deep learning models rely on data distribution in the feature space, while ECG signals vary considerably from person to person, making it difficult to reliably apply models trained on particular ECG datasets in practice. While existing reviews primarily focus on deep learning algorithms, this study takes a comprehensive approach, considering various factors throughout the entire deep learning workflow for ECG arrhythmia classification. The release of a large, labeled, multinational, 12-lead ECG dataset as part of the 2020 PhysioNet Challenge\cite{b5} presents a unique opportunity to advance multi-class cardiac abnormality detection.

This work addresses the problem using a data augmentation technique on a deep neural network. The architecture includes a squeeze-and-excitation (SE) block on the ResNet-18 model to acknowledge the importance of the spatial relationship between ECG channels\cite{b6}. Data augmentation by signal transforms is a valuable technique for analyzing ECGs; however, acquiring a large and diverse dataset of ECG signals can be challenging. By applying signal transforms to existing ECG signals, it is possible to generate new synthetic samples that closely resemble real-world variations. This augmented dataset enables more robust training of deep neural networks, improving performance in ECG signal classification. By employing data augmentation through signal transforms, researchers and healthcare professionals can unlock deeper perspectives from ECG data and enhance patient care. Sinusoidal Time--Amplitude Resampling (STAR) is presented as a simple beat-wise augmentation that applies controlled time warping and matched amplitude scaling per R--R segment, preserving P--QRS--T morphology and avoiding the global gain drift typical of Multiply--Triangle (see Sec.~\ref{sec:star}).

\section{\textbf{Related Works}}
In recent times, the convergence of medical treatment and artificial intelligence (AI) has attracted increasing attention among scholars. This multidisciplinary realm has become a pivotal focus for numerous corporate entities, academic institutions, and research centers, catalyzed by its substantial scientific and economic promise. The domain of deep learning centers on extracting, forecasting, and making decisions using a foundational dataset—referred to as training data—with the intention of revealing latent information and patterns. Over the past decades, a wide array of machine learning models has been conceived and refined to enhance the precision of diverse learning tasks. These models span decision trees, K-Nearest Neighbors, ensemble methods, and various forms of deep learning models.

In an effort to establish robust and dependable cardiac monitoring systems, R. Smisek introduced an SVM-based classifier capable of discerning various forms of atrial and ventricular activity, heartbeats, and, ultimately, ECG records. This multi-tiered approach combines a support vector machine (SVM), a decision tree, and threshold-based rules, aiming to enhance diagnostic accuracy, reduce labor-intensive expert ECG scoring, and ultimately expedite timely and effective treatment \cite{b17}. 
In recent years, a diverse array of deep learning models—encompassing the multi-layer perceptron, CNNs, LSTM, and more intricate architectures such as ResNet-18—has been developed and refined, assuming a pivotal role within this domain. While originally devised for image processing, ResNet-18 can be adapted for signal-processing tasks. Moreover, tailored architectures optimized for specific signal data types also exist. Deep learning techniques have been extensively scrutinized for anomalies in cardiac function, such as arrhythmia diagnosis using ECG signals, showing promising potential for integration into clinical applications. Nonetheless, further research is warranted before incorporating DL pipelines for dependable clinical ECG classification. These DL models comprise multi-tiered or multi-layered structures, with each tier or layer functioning as a feature extractor adept at acquiring insights into signal characteristics \cite{b9}.

Given the inherent nature of major feature-extraction components within neural networks, the DL classification models explored in the selected investigations primarily fall into the following categories: convolutional neural networks (CNNs); recurrent neural networks (RNNs), including long short-term memory (LSTM) and bidirectional LSTM (BiLSTM); transformer architectures; “hybrid” configurations that combine different DL models; and “others,” representing less commonly employed models such as restricted Boltzmann machines and deep-belief networks. A comprehensive analysis of these DL models in the context of ECG arrhythmia classification is as follows.

CNNs that renowned in image classification, signal analysis, and natural language processing, also find utility in the ECG domain \cite{rtdetrMaritime2025}. For example, Kiranyaz proposed an adaptive 1D CNN for ECG classification and anomaly detection, obviating the need for manual feature extraction \cite{b10}. Y. Jin harnessed 1D CNNs for the classification of various heart diseases, including scenarios with limited dataset sizes, employing few-shot learning techniques. In contrast, 2D CNNs primarily handle image-like inputs such as ECG signal spectrograms and scalograms \cite{b11}. A. Mostayed devised a model in which 2D scalograms are generated from 1D ECG signals via continuous wavelet transform \cite{b12}.

With a focus on temporal correlations within feature sequences, RNNs emerge as a class of DL structures well-suited for time-series inputs like ECG signals. Because ECG signals inherently exhibit temporal dependencies, capturing their temporal features holds significant promise for accurate categorization. In traditional RNNs, the hidden layer’s state at a given moment relies not only on the current input but also on prior temporal instances, rendering them adept at capturing hidden temporal characteristics within ECG features \cite{b13}. Bidirectional LSTM (BiLSTM), a specialized form of LSTM, incorporates two LSTM units that traverse the input sequence forward and backward, respectively \cite{b14}. Additionally, the transformer architecture—comprising attention mechanisms and fully connected layers—has gained prominence \cite{b15}. G. Yan’s work harnesses the transformer encoder for heartbeat classification using ECG signals, with concatenated RR intervals enhancing the final classification \cite{b16}. Further research avenues and prospects include leveraging diverse ECG databases for training and testing, innovative denoising and data augmentation techniques, novel integrated DL architectures, and a deeper exploration of inter-patient paradigms. These endeavors aim to establish trustworthy DL-based classification frameworks and facilitate their integration into real-world clinical scenarios. For instance, Z. Xiong devised RhythmNet, a 21-layer convolutional recurrent neural network for classifying cardiac rhythms from ECG recordings, presenting potential clinical applications for AF diagnosis and self-monitoring, and achieving an accuracy of 82\% \cite{b19}.

\section{\textbf{Background}}

\subsection{\textbf{ECG Signals}}
This research focuses on the multi-label classification of cardiac abnormalities using 12-lead ECG recordings that exhibit discrepancies in both sampling frequency and duration. These recordings are a standard instrument in clinical environments, used to diagnose conditions such as arrhythmias, atrial fibrillation, and coronary occlusion~\cite{b20}.

\textit{Data sources} In line with the PhysioNet/Computing in Cardiology (CinC) 2020 Challenge, the training–evaluation data are drawn from multiple clinical repositories with imbalanced acquisition settings. Specifically, the combined collection comprises over 90{,}000 standard 12-lead ECGs sourced from Chinese and international healthcare systems (including CPSC/CPSC-Extra, SPH, G12EC, Chapman–Shaoxing/Ningbo, and UMich), with sampling frequencies typically ranging from \SIrange{250}{1000}{Hz} and record durations between $\sim$\SI{6}{s} and \SI{60}{s}~\cite{cinc2020-overview,b5}. This diversity reflects real-world deployment conditions (different devices, paper vs.\ digital workflows, and patient populations), which is crucial for developing generalizable ECG classifiers.

Each ECG record in these collections is accompanied by basic demographics (age, sex) and one or more clinical labels indicating rhythm or morphological abnormalities. Multi-label annotation is common, mirroring clinical reality, where concurrent diagnoses can co-occur in the same patient record. Variability in sampling rates and durations can directly affect frequency content (e.g., representation of P-, QRS-, and T-wave harmonics; i.e., the P–QRS–T morphology) and the effective context available to the model, motivating careful pre-processing and model design to ensure comparability across inputs~\cite{cinc2020-overview}.

\subsection{\textbf{Deep Learning}}
Deep learning provides end-to-end representation learning that can automatically extract hierarchical features from raw waveform inputs without handcrafted engineering. For ECGs, this is particularly advantageous because diagnostically relevant cues span scales—from millisecond QRS upstrokes to multi-second rhythm patterns. Convolutional encoders have demonstrated strong performance on ECG classification by learning local morphological filters and progressively aggregating temporal context~\cite{b10}. Compared with traditional pipelines that rely on predefined features, deep networks better adapt to cross-dataset shifts (e.g., different sampling rates or noise characteristics), provided that appropriate normalization, data balancing, and augmentation are employed.

\subsection{\textbf{Convolutional Neural Network}}
Convolutional neural networks (CNNs) have been widely employed for 2D/3D vision tasks and are equally effective for one-dimensional biosignals. In 1D form, CNNs convolve learnable kernels along time, capturing local morphology (e.g., QRS slope, ST segment, T-wave symmetry), while pooling layers downsample and provide robustness to small temporal misalignments. As the signal propagates through deeper layers, learned features transition from primitive edge-like patterns to class-specific motifs, which a final classifier can leverage. The mathematical operation for a 1D convolution at time~$t$ is
\begin{equation}\label{eq7}
C_t \;=\; f\!\Big(\sum_{j=-(l-1)/2}^{(l-1)/2} \omega_j \, X_{t-j} + b\Big), \qquad t=1,\dots,T.
\end{equation}
where $\omega$ is a filter of length~$l$, $b$ a bias, $f(\cdot)$ a nonlinearity, and $X$ the input sequence. A sliding-window diagram illustrates this intuition (omitted for brevity). Backpropagation is then used to optimize all parameters jointly by minimizing a task-specific loss over labeled ECGs.

Convolutional architectures used here follow the standard pattern of stacked temporal convolutions and pooling, culminating in fully connected layers for multi-label outputs. The convolutional layers detect local features by scanning multiple kernels across the signal, while pooling reduces redundancy and provides a degree of invariance to beat-to-beat variation. As depth increases, learned features become more discriminative for the final classification task (see Eq.~\ref{eq7})~\cite{b22,b23}. Backpropagation iteratively adjusts the filters and classifier to minimize loss and improve predictive performance.

Generalization across devices, care settings, and demographics benefits from targeted data augmentation. For time-series ECGs, useful families include amplitude/time scaling, random cropping, local warps, noise injections, and frequency-domain perturbations; recent surveys provide taxonomies and guidelines for selecting augmentations that preserve diagnostic semantics while enriching variability~\cite{iwana2021augment}. In multi-label scenarios, augmentation can also mitigate class imbalance without oversimplifying minority phenotypes.

\section{\textbf{Methods}}\label{sec:methods}

The primary objective of this study is to develop a model that \emph{reliably} classifies 12-lead ECG recordings into \textbf{14} diagnostic classes while systematically quantifying the impact of data augmentation on multi-label performance. Because clinical ECGs often exhibit \emph{co-occurring} conditions, the learning problem is explicitly framed as multi-label classification with independent decision functions per class.

\subsection{\textbf{Data sources}}
\label{subsec:data-sources}

A five-source, multi-institutional corpus of standard 12-lead ECGs was assembled by combining the public training portion of the George B.\ Moody PhysioNet/Computing in Cardiology (CinC) Challenge~2021 with an independent release from Shandong Provincial Hospital (SPH). Following the Challenge overview, the CinC collection is treated as four consolidated sources, including Chapman–Shaoxing\,+\,Ningbo, CPSC\,+\,CPSC-Extra, PTB\,+\,PTB-XL, and G12EC. SPH also serves as an independent fifth source. Table~\ref{tab:data_sources} reports, for each source, the acquisition site(s), country, nominal sampling rates, record lengths, and whether labels follow SNOMED CT or the AHA/ACC/HRS statement\,+\,modifier convention. As summarized there, the pooled corpus retains \num{103438} ECGs after label filtering spanning heterogeneous devices, workflows, and patient populations. This heterogeneity plays an essential role in intentionally stressing generalization.

\paragraph*{Chapman–Shaoxing \& Ningbo.}
This combined Chinese cohort (10-second, 500\,Hz 12-lead ECGs) contributes the largest single-site share of records among the CinC sources. Labels originate in SNOMED~CT and cover common rhythm and morphological findings. Its scale and uniform sampling facilitate stable model calibration while still reflecting real clinical diversity (see Table~\ref{tab:data_sources}).

\paragraph*{CPSC \& CPSC-Extra.}
These sets derive from the China Physiological Signal Challenge releases and include 12-lead ECGs with variable durations (approximately 6--144\,s) at 500\,Hz. Labels follow the CinC SNOMED coding. Their broader duration range enriches rhythm context and exposes the pipeline to variable temporal coverage (Table~\ref{tab:data_sources}).

\paragraph*{G12EC.}
The Georgia 12-Lead ECG Challenge database (Emory University Hospital) provides U.S.\ clinical ECGs (5--10\,s; 500\,Hz) labeled in SNOMED. By design, it introduces a different patient mix and device ecology, aiding cross-system robustness (Table~\ref{tab:data_sources}).

\paragraph*{PTB \& PTB-XL.}
The PTB family (Germany/EU) contributes 10--120\,s records sampled at 500 or 1000\,Hz with SNOMED labels. PTB-XL in particular supplies standardized, large-scale clinical ECGs that broaden morphology and rhythm coverage and stress-test downsampling/harmonization (Table~\ref{tab:data_sources}).

\paragraph*{SPH.}
The SPH release (China) contains routine 12-lead clinical ECGs (10--60\,s; 500\,Hz) labeled using the AHA/ACC/HRS interpretive statement scheme with optional modifiers. SPH acts as the only non-CinC source and is therefore critical for evaluating label harmonization and transfer across coding systems (Table~\ref{tab:data_sources}).

\medskip
Table~\ref{tab:data_sources} enumerates acquisition characteristics and label standards per source, making explicit the heterogeneity (sampling rate, duration, and coding) that the preprocessing must reconcile. Table~\ref{tab:demo} then summarizes demographics, age and sex distributions. Together, these tables motivate the harmonization pipeline and the source-aware evaluation protocol.

\medskip
CinC sources use SNOMED~CT labels, whereas SPH uses AHA statements with optional modifiers. A cardiologist-guided mapping converts each SPH statement\,+\,modifier into the corresponding SNOMED concept from the Challenge scoring list to ensure parity across sources. In this harmonization, SPH \textit{Normal ECG} is assigned \(-1\) to preserve records during filtering. All \textit{prolonged PR} variants merge under \textit{first-degree atrioventricular block}, and statement–modifier phenotypes such as \textit{atrial fibrillation} as well as \textit{premature atrial contraction} collapse to single SNOMED labels. Table~\ref{tab:label-counts} reports post-mapping diagnosis counts per source, illustrating label prevalence and cross-source balance achieved by the mapping.

\medskip
\noindent\textit{Multiplicity of labels}
Clinical ECGs often carry multiple co-occurring diagnoses. Figure~\ref{fig:multilabel} visualizes, by source, the proportion of records with 1, 2, 3, or $\geq$4 labels (normalized to percent). This highlights the dataset-specific co-occurrence structure that the multi-label model and stratification need to respect.

\medskip
Regarding the preprocessing stage, all ECGs are harmonized to a consistent $12\times 4096$ tensor at a minimum effective sampling of $\geq\!250$\,Hz. Sequences shorter than 4096 samples are \emph{randomly zero-padded} to preserve phase diversity, and longer recordings are \emph{randomly clipped} with a 4096-sample window to reduce onset/offset bias across rhythms. Optional per-lead z-score normalization was explored in ablations. Demographics are standardized, age min–max to $[0,1]$ and sex one-hot with explicit missingness masks, so the network can learn to ignore absent covariates. After the SNOMED\,$\leftrightarrow$\,AHA mapping, each record is encoded as a 14-dimensional multi-hot vector $\mathbf{y}\in\{0,1\}^{14}$; ambiguous/conflicting-label records are excluded from training folds. These steps ensure that inputs are comparable across devices and sites while preserving clinically meaningful temporal and morphological content.

\begin{table*}[t]
\caption{Overview of data sources used in this study}
\label{tab:data_sources}
\centering
\footnotesize
\setlength{\tabcolsep}{12pt}
\renewcommand{\arraystretch}{2.0}
\small
\begin{adjustbox}{max width=\textwidth}
\begin{tabular}{l l l r r r l l l}
\toprule
\textbf{Source} & \textbf{Country} & \textbf{Location} &
\textbf{Patients} & \textbf{ECGs} & \textbf{Included} &
\textbf{Sampling (Hz)} & \textbf{Length (s)} & \textbf{Labeling} \\
\midrule
Chapman--Shaoxing \& Ningbo & China &
Shaoxing People's Hospital; Ningbo First Hospital & 45\,152 & 45\,152 & 43\,814 & 500 & 10 & SNOMED \\
CPSC \& CPSC-Extra          & China & 11 unnamed hospitals & -- & 10\,330 & 6\,110  & 500 & 6--144 & SNOMED \\
G12EC                       & USA   & Emory University Hospital & 15\,738 & 10\,344 & 8\,892  & 500 & 5--10 & SNOMED \\
PTB \& PTB-XL               & DE/EU & PTB / Technische Bundesanstalt & 19\,147 & 22\,353 & 21\,348 & 500, 1000 & 10--120 & SNOMED \\
SPH                         & China & Shandong Provincial Hospital & 24\,666 & 25\,770 & 23\,274 & 500 & 10--60 & AHA \\
\bottomrule
\end{tabular}
\end{adjustbox}
\end{table*}

\subsubsection*{\textbf{Demographics}}
Age/sex meta-data (mean\,\,$\pm$\,SD, min–max, and undefined counts U) are shown in Table~\ref{tab:demo}. The overall cohort skews slightly male (56/44).

\begin{table*}[t]
\caption{Demographic information per source.}
\label{tab:demo}
\centering
\setlength{\tabcolsep}{10pt}
\begin{tabular*}{\textwidth}{@{\extracolsep{\fill}} l r@{\,}l r r r c @{}}
\toprule
\multirow{2}{*}{\textbf{Source}} & \multicolumn{3}{c}{\textbf{Age}} & \multicolumn{2}{c}{\textbf{Sex (\%)}} \\
\cmidrule(lr){2-4}\cmidrule(l){5-6}
& \multicolumn{1}{c}{Mean} & \multicolumn{1}{c}{$\pm$SD} & \multicolumn{1}{c}{Min–Max / U} & \multicolumn{1}{c}{M} & \multicolumn{1}{c}{F} \\
\midrule
Chapman–Shaoxing \& Ningbo & 58.21 & 19.62 & 0–89 / 44  & 56 & 43 \\
CPSC \& CPSC-Extra          & 62.69 & 18.67 & 1–104 / 8  & 57 & 43 \\
G12EC                       & 60.82 & 15.51 & 14–89 / 65 & 54 & 46 \\
PTB \& PTB-XL               & 59.57 & 17.01 & 2–95 / 92  & 52 & 48 \\
SPH                         & 49.66 & 15.51 & 18–95 / 0  & 57 & 43 \\
\midrule
\textbf{Final dataset}      & 57.05 & 18.32 & 0–104 / 209 & 56 & 44 \\
\bottomrule
\end{tabular*}
\vspace{-1ex}
\begin{minipage}{\textwidth}
\footnotesize
\vspace{1ex}
U = Undefined. Chapman–Shaoxing \& Ningbo has 14 ECGs with missing sex values reported in the source; here U counts refer to age.
\end{minipage}
\end{table*}

CinC sources are annotated with \emph{SNOMED~CT} codes, whereas SPH uses the \emph{AHA/ACC/HRS} interpretive statement convention with optional modifiers. A cardiologist-reviewed mapping converts AHA statements\,+\,modifiers to the corresponding SNOMED concepts used by the Challenge scoring list, so labels are comparable across sources. Within this harmonization, \textit{Normal ECG} in SPH is assigned $-1$ to preserve records during filtering. \textit{Prolonged PR} variants merge under \textit{first-degree atrioventricular block}, as well as statement–modifier phenotypes such as \textit{atrial fibrillation} and \textit{premature atrial contraction} are collapsed to single SNOMED labels to maintain parity with CinC. Table~\ref{tab:label-counts} reports representative diagnosis counts per source after mapping.

\begin{table*}[t]
\centering
\caption{Selected diagnosis counts per source after mapping (subset shown; a full table can be placed in the appendix).}
\label{tab:label-counts}
\setlength{\tabcolsep}{10pt}
\renewcommand{\arraystretch}{1.15}
\begin{adjustbox}{max width=\textwidth}
\begin{tabular}{@{}l l
                S[table-format=6.0]
                S[table-format=5.0]
                S[table-format=5.0]
                S[table-format=5.0]
                S[table-format=5.0]
                S[table-format=6.0]@{}}
\toprule
\textbf{Diagnosis} & \textbf{Standard} &
\multicolumn{1}{c}{\textbf{Chap.\,\&\,Ningbo}} &
\multicolumn{1}{c}{\textbf{CPSC\,+\,Extra}} &
\multicolumn{1}{c}{\textbf{G12EC}} &
\multicolumn{1}{c}{\textbf{PTB\,+\,XL}} &
\multicolumn{1}{c}{\textbf{SPH}} &
\multicolumn{1}{c}{\textbf{Total}}\\
\midrule
first-degree AV block            & SNOMED/AHA$^{\dagger}$ & 828  &    &  769 & 1012 & 238 & 4039 \\
atrial fibrillation              & SNOMED/AHA             & 1780 & 1374 &  570 & 1529 & 675 & 5928 \\
atrial flutter                   & SNOMED/AHA             & 8060 &   51 &  186 &   74 &  99 & 8473 \\
incomplete RBBB                  & SNOMED/AHA             & 3060 &  407 & 1118 & 1259 &    & 3116 \\
LAFB                             & SNOMED/AHA             &  102 &  180 & 1626 &  154 &    & 2340 \\
left axis deviation              & SNOMED/AHA             & 1355 &  940 &  514 &  138 &    & 3769 \\
left bundle branch block         & SNOMED/AHA             &  274 &     &  321 &  536 &  84 & 1365 \\
low QRS voltages                 & SNOMED/AHA             &  125 &     &  240 &  382 & 122 & 1921 \\
premature atrial contraction     & SNOMED/AHA             & 1312 &  689 &  639 &  398 & 539 & 3577 \\
right axis deviation             & SNOMED/AHA             &  853 &    1 &   83 &  343 & 161 & 1441 \\
right bundle branch block        & SNOMED/AHA             &      & 1858 &  542 &      & 710 & 3759 \\
sinus arrhythmia                 & SNOMED/AHA             & 2550 &     &  772 & 1553 &     & 5341 \\
sinus bradycardia                & SNOMED/AHA             & 1677 &   45 &      & 2711 &     & 21629 \\
sinus rhythm                     & SNOMED/AHA$^{\ddagger}$& 8125 &     & 18172&      &16858& 45829 \\
sinus tachycardia                & SNOMED/AHA             & 3480 &     &  1261&  827 & 725 & 10371 \\
T wave abnormal                  & SNOMED/AHA             &  743 &   22 &  2306& 2345 &2042 & 13758 \\
T wave inversion                 & SNOMED/AHA             &      &    5 &   812&  294 & 176 &  4164 \\
\bottomrule
\end{tabular}
\end{adjustbox}

\vspace{-1ex}
\begin{minipage}{\textwidth}
\footnotesize
\vspace{2.5ex}
$^{\dagger}$\,All ``prolonged PR'' statements in SPH merged into first-degree AV block.\;
$^{\ddagger}$\,SR imputed for abnormal SPH ECGs using a logistic model trained on CinC multi-labels; normal SPH ECGs set to SR by default (see text).
\end{minipage}
\end{table*}

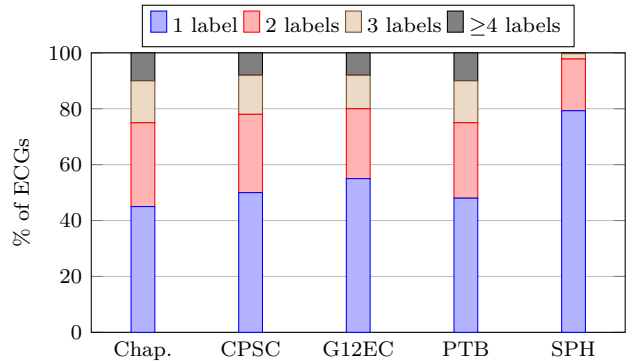
\begin{figure}[htbp!]
\centering
\begin{tikzpicture}
\begin{axis}[
    width=0.98\columnwidth,
    height=5.3cm,
    ybar stacked,
    ymin=0, ymax=100,
    bar width=9pt,
    xtick=data,
    xticklabels={Chap., CPSC, G12EC, PTB, SPH},
    legend style={at={(0.5,1.02)},anchor=south,legend columns=5, font=\footnotesize},
    ylabel={\% of ECGs},
    ymajorgrids,
    enlarge x limits=0.12,
    tick label style={font=\footnotesize},
    label style={font=\footnotesize}
]
\addplot coordinates {(1,45) (2,50) (3,55) (4,48) (5,79.3)}; 
\addplot coordinates {(1,30) (2,28) (3,25) (4,27) (5,18.5)}; 
\addplot coordinates {(1,15) (2,14) (3,12) (4,15) (5,1.9)};  
\addplot coordinates {(1,10) (2,8) (3,8) (4,10) (5,0.3)};   
\legend{1 label, 2 labels, 3 labels, $\ge$4 labels}
\end{axis}
\end{tikzpicture}
\caption{Normalized distribution of label multiplicity per source (percent). Exact SPH percentages are shown; CinC bars can be updated with precise values if desired.}
\label{fig:multilabel}
\end{figure}

Considering the details of the data, for SPH specifically, \SI{79.3}{\percent} have one label, \SI{18.5}{\percent} two, \SI{1.9}{\percent} three, and \SI{0.3}{\percent} four or five labels. Figure~\ref{fig:multilabel} shows normalized per-source distributions (percent), which differ from raw counts to improve readability in two columns.

\subsection{\textbf{Data pre-processing}}

In the context of signal harmonization, all recordings are resampled, preferably, to a minimum effective sampling rate of $\geq\!250$\,Hz to ensure comparable temporal resolution across sources. Each 12-lead trace is then represented as a fixed window $\mathbb{R}^{12\times 4096}$. The sequences shorter than 4096 samples are \emph{randomly zero-padded} to preserve phase diversity across beats. Otherwise, the longer sequences are \emph{randomly clipped} to a 4096-sample window to avoid selection bias toward any specific onset. Furthermore, per-lead z-score normalization is evaluated in ablation to reduce inter-device amplitude scale differences (not used in the final model). Age and sex are standardized as auxiliary covariates. Age is min–max scaled to $[0,1]$ and sex is one-hot encoded with an explicit missingness mask so the network can learn to ignore absent metadata at inference. These choices follow common practice for imbalanced multi-institutional 12-lead ECG corpora with variable sampling and duration~\cite{cinc2020-overview,ptbxl-sndata-2020}.

\textbf{Label preparation.} Source-provided diagnostic statements are mapped to a unified 14-class ontology. Records with ambiguous or conflicting labels are excluded from training folds. Multi-label targets are stored as binary vectors $\mathbf{y}\in\{0,1\}^{14}$ and used with independent sigmoid outputs during training and evaluation.

\subsection{\textbf{Model}}
\begin{figure*}[htbp]
  \centering
  \subfloat[Backbone with SE blocks\label{fig:arch}]{%
    \includegraphics[width=0.48\linewidth]{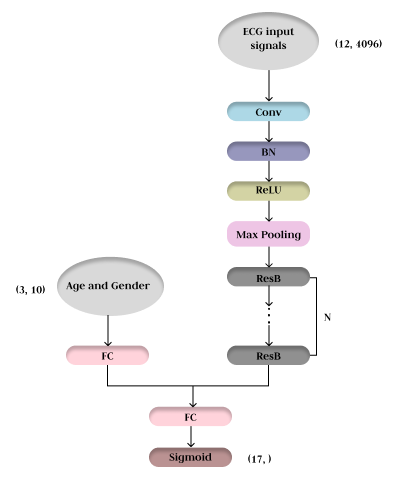}
  }\hfill
  \subfloat[Residual block with SE module\label{fig:seblock}]{%
    \includegraphics[width=0.48\linewidth]{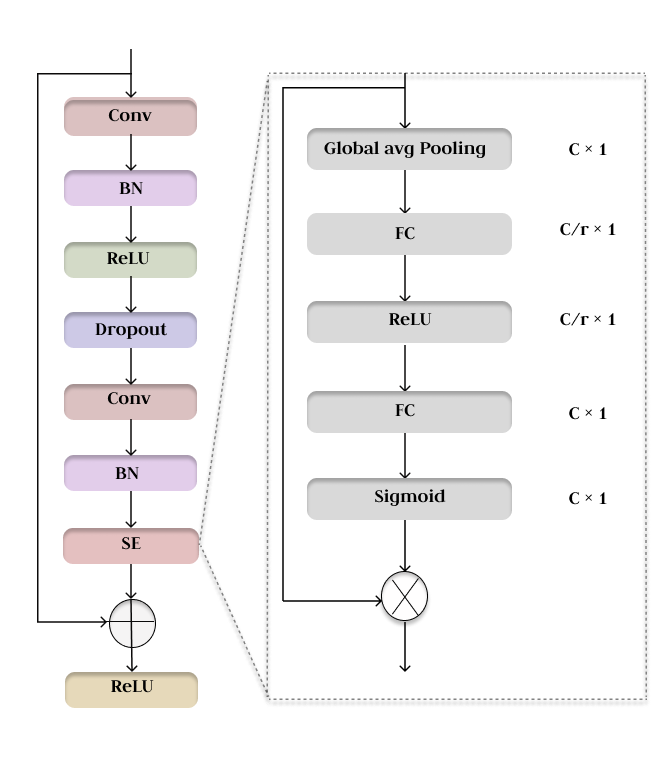}
  }
  \caption{SE–ResNet-18 classifier and its SE-augmented residual block shown side by side for reference.}
  \label{fig:model}
\end{figure*}

The classifier is a \textbf{1D SE–ResNet-18} for 12-lead ECG waveforms (overall layout in Fig.~\ref{fig:model}\,\subref{fig:arch}). Each residual block applies temporal convolutions along the time axis to produce lead-wise feature maps, following standard ECG CNN and ResNet practice. To enhance cross-lead saliency, every block includes a squeeze-and-excitation (SE) unit that performs global temporal pooling and lightweight channel re-weighting before the skip addition (module detail in Fig.~\ref{fig:model}\,\subref{fig:seblock}), a mechanism known to improve representational power with modest overhead~\cite{hu2018senet}. 

The network begins with a temporal convolutional stem and proceeds through eight residual blocks with progressive downsampling (Fig.~\ref{fig:model}\,\subref{fig:arch}). Demographic variables are projected with a small fully connected layer and concatenated with the globally pooled features before the final classifier. Since diagnoses are not mutually exclusive, the output head uses independent sigmoid activations, one per class, and is trained with class-weighted binary cross-entropy to address label imbalance and avoid softmax competition. 

Unless noted otherwise, optimization uses NAdam with a cosine-annealed learning rate schedule and early stopping on validation AUROC, with per-class loss weights set inversely to prevalence.

\subsection{\textbf{Data Augmentation}}
To improve generalization while preserving diagnostic morphology, a morphology-preserving, beat-aligned augmentation is adopted and compared with standard time-domain transforms. Detailed settings and activation probabilities for comparator policies (Multiply--Triangle, temporal shift, light Gaussian noise, rare lead dropout) are reported once in Sec.~\ref{sec:experiments} and Tables~\ref{tab:band_params}--\ref{tab:final_params} to avoid duplication.

\vspace{2mm}
\noindent\textbf{Sinusoidal Time--Amplitude Resampling (STAR)}\label{sec:star}\enspace

\noindent
\textbf{Motivation.}
Common ECG augmentations such as random cropping, strong global time warps, or stacked triangles can drop or shift R-peaks, truncate intervals, or exaggerate noise. \emph{STAR} avoids these pitfalls by operating \textbf{beat-wise} between consecutive R-peaks, introducing smooth, parameterized variability while keeping clinical morphology intact.

\vspace{1.2mm}
\noindent
\textbf{Method.}
Let $R=\{R_1,\dots,R_K\}$ be the R-peak indices on a reference lead, partitioning the trace into $K{-}1$ consecutive \textbf{R--R} segments. For segment $i\!\in\!\{1,\dots,K{-}1\}$ (from $R_i$ to $R_{i+1}{-}1$) and lead $\ell$:
\begin{enumerate}\itemsep3pt
  \item Define $x_\ell^{(i)} = X_\ell[R_i:R_{i+1})$ with length $N_i=R_{i+1}-R_i$.
  \item Draw a coefficient $c_i$ from a sinusoidal schedule across segments:
  \[
  c_i \;=\; A_3 + (A_2-A_3)\,\frac{\sin\!\Bigl(2\pi\,\frac{i-1}{K-1}+\phi\Bigr)+1}{2}, \quad A_2>A_3.
  \]
  \item \textbf{Time warp:} resample $x_\ell^{(i)}$ via monotone interpolation to
  $m_i=\max\!\bigl(1,\lfloor c_i\,N_i\rfloor\bigr)$ samples.
  \item \textbf{Amplitude scale:} multiply the resampled segment by the same $c_i$.
\end{enumerate}
Finally, concatenate the warped segments in order and reattach the exact head $X_\ell[0:R_1)$ and tail $X_\ell[R_K:T)$, padding or trimming to $T$ if needed. A complete pseudo-code is provided in \textbf{Algorithm~\ref{alg:star}}.

\vspace{1.2mm}
\noindent
\textbf{Why STAR preserves information (concise).}
Segmentation is strictly \textbf{R--R}; the head and tail are reattached \emph{unchanged}; and per-segment warping uses monotone interpolation with linear amplitude scaling. Peaks and clinically relevant intervals are therefore retained, while controlled variability is added.

\paragraph{\textbf{Theoretical property (invertibility)}}
Let $X\in\mathbb{R}^{L\times T}$ be a multi-lead ECG and let $R=\{R_1,\dots,R_K\}$ denote R-peak indices on a reference lead that induce $K-1$ consecutive R--R segments. The proposed STAR transform on each lead is the composition
\[
\mathcal{T}_{\mathrm{STAR}}
\;=\;
\mathcal{H}\circ\mathcal{C}\circ
\bigl(\,\mathcal{S}_{\mathbf{c}}\circ \mathcal{W}_{\mathbf{c}}\circ \mathcal{E}_{\mathbf{S}}\,\bigr),
\]
where $\mathcal{E}_{\mathbf{S}}$ equalizes segment lengths to prescribed integers $\mathbf{S}=(S_1,\dots,S_{K-1})$ via monotone interpolation, $\mathcal{W}_{\mathbf{c}}$ resamples each equalized segment by a positive factor $c_i\in[A_3,A_2]$ using monotone interpolation, $\mathcal{S}_{\mathbf{c}}$ scales amplitudes by the same $c_i$, $\mathcal{C}$ concatenates the processed segments, and $\mathcal{H}$ reattaches the unmodified head $X[0\!:\!R_1)$ and tail $X[R_K\!:\!T)$ with a length–preserving body projection.


\begin{proposition}[STAR is invertible under mild conditions]\label{prop:star-invertible}
Assume the following:
\begin{enumerate}[label=(\roman*),leftmargin=*,itemsep=2pt]
  \item the R-peak set $R$ is correct and stored;
  \item the interpolation operators used by $\mathcal{E}_{\mathbf{S}}$ and $\mathcal{W}_{\mathbf{c}}$ are monotone time warps with known sampling grids;
  \item per-segment amplitude factors satisfy $c_i>0$ and are stored; and
  \item each segment is bandlimited below the effective Nyquist after resampling.
\end{enumerate}
Then $\mathcal{T}_{\mathrm{STAR}}$ is invertible on the class of piecewise bandlimited signals segmented by $R$, with an inverse
\begin{equation*}
\mathcal{T}_{\mathrm{STAR}}^{-1}
=
\bigl(\mathcal{E}_{\mathbf{S}}\bigr)^{-1}
\circ
\bigl(\mathcal{W}_{\mathbf{c}}\bigr)^{-1}
\circ
\bigl(\mathcal{S}_{\mathbf{c}}\bigr)^{-1}
\circ
\mathcal{C}^{-1}
\circ
\mathcal{H}^{-1},
\end{equation*}
and the reconstruction error is bounded by the interpolation error.
\end{proposition}

$\mathcal{S}_{\mathbf{c}}$ is trivially invertible per segment since $c_i>0$ and $(\mathcal{S}_{\mathbf{c}})^{-1}$ multiplies by $1/c_i$.
Given stored sampling grids, each monotone resampling $\mathcal{W}_{\mathbf{c}}$ admits the left inverse that interpolates back to the original grid; likewise for $\mathcal{E}_{\mathbf{S}}$.
$\mathcal{C}$ and $\mathcal{H}$ are bookkeeping operators that concatenate segments and reattach the unchanged head and tail; their inverses recover segment boundaries from the stored metadata $(R,\mathbf{S},\mathbf{c})$.
Under bandlimitedness and Nyquist, interpolation is a stable isomorphism up to a small bounded error. Therefore, the composite inverse exists, and the overall reconstruction error is dominated by interpolation error.\\[1mm]
Under mild assumptions on segmentation and resampling, STAR is invertible, and its frequency behavior can be formalized. A constructive proof with an algorithmic inverse is given in Proposition~\ref{prop:star-invertible} and Appendix~\ref{app:invertibility}.

\vspace{-2mm}
\noindent
\textbf{Intuition.}
In plain terms, STAR gently stretches or compresses each beat and scales it up or down by the same factor—but only within consecutive R--R intervals. This keeps P--QRS--T ordering intact and leaves the trace head and tail untouched, so diagnostic landmarks remain where clinicians expect them, while the model sees realistic variability.

\vspace{10mm}
\noindent
\textbf{Recipe (inputs, steps, hyperparameters).}
\begin{tcolorbox}[title=\textbf{STAR Recipe}]
\textbf{Inputs.} Multi-lead ECG $X\!\in\!\mathbb{R}^{L\times T}$; detected R-peaks $R=\{R_i\}_{i=1}^K$ on a reference lead.\\
\textbf{Steps.} (1) Partition into R--R segments; keep $X[0\!:\!R_1)$ and $X[R_K\!:\!T)$ unchanged. (2) For segment $i$, compute $c_i$ via the sinusoidal schedule. (3) Resample segment $i$ to $m_i=\max(1,\lfloor c_i N_i\rfloor)$ (monotone interpolation). (4) Multiply by $c_i$. (5) Concatenate all segments, then reattach head/tail; pad/trim to length $T$.\\
\textbf{Hyperparameters.} Bounds $(A_2,A_3)$ with $A_2\!>\!A_3$; phase $\phi$; number of sinusoid periods $n$ (default $n{=}1$); activation probability $p$.\\
\textbf{Suggested ranges.} $A_2\!\in\![1.2,2.0]$,\; $A_3\!\in\![0.4,0.9]$,\; $\phi\!\in\!\{0,\pi/2,\pi\}$,\; $n\!\in\!\{1,2\}$,\; $p\!\in\!\{0.25,0.5,0.75\}$.\\
\textbf{Final settings (this study).} $(A_2,A_3,\phi,n)=(1.6,\,0.6,\,0,\,1)$,\; $p=0.5$.
\end{tcolorbox}

\vspace{0.8mm}
\noindent
\begin{figure}[t]
  \centering
  \includegraphics[width=0.92\columnwidth]{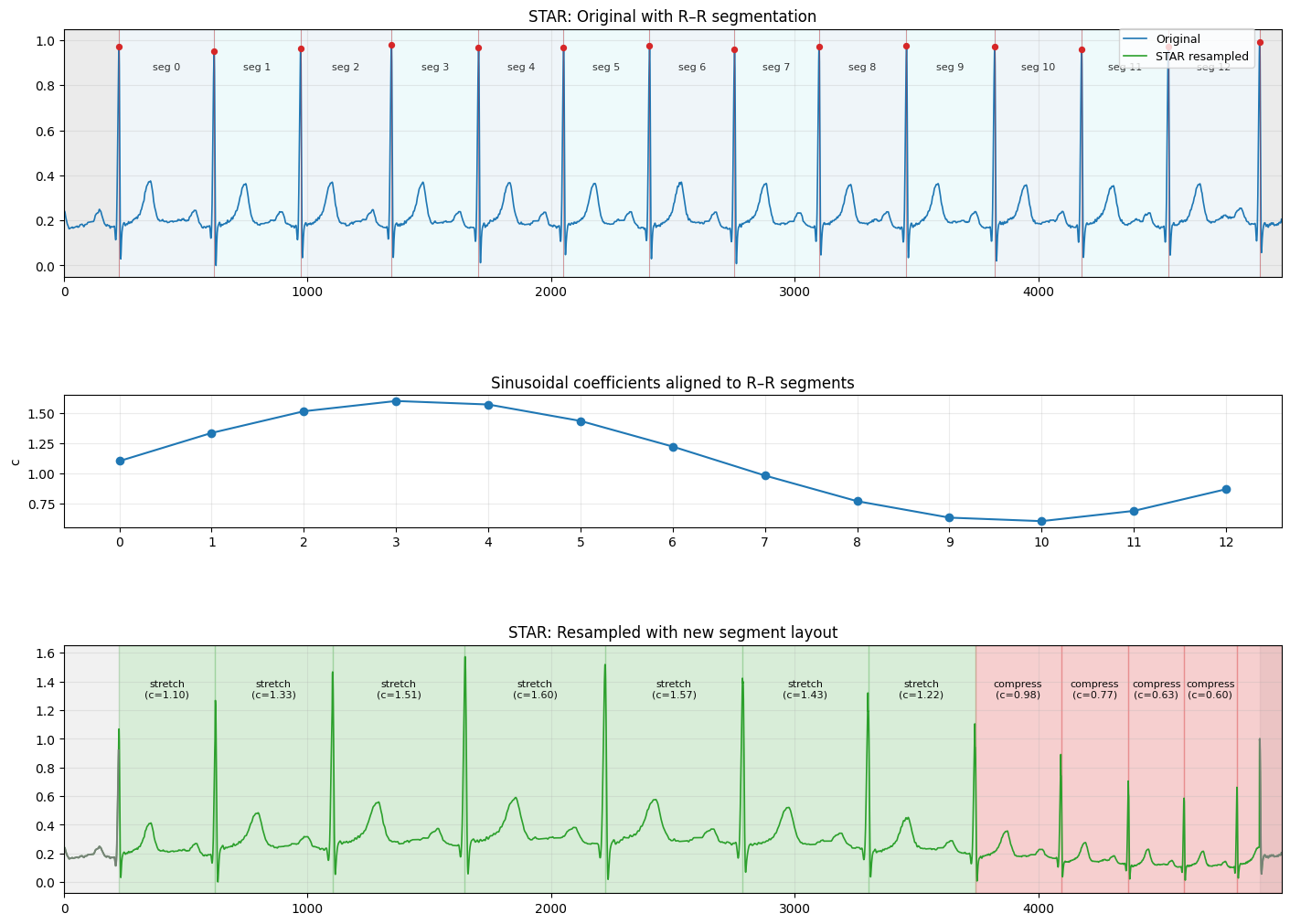}
  \caption{STAR schematic. Top: original ECG with detected R-peaks. Middle: partition into consecutive R--R segments. Bottom: per-segment time resampling and amplitude scaling by sinusoidal coefficients, followed by exact head/tail reattachment; P--QRS--T ordering preserved.}
  \label{fig:star-schematic}
\end{figure}

\begin{algorithm}[htbp]
\caption{STAR Resampling (Sinusoidal Time--Amplitude Resampling)}
\label{alg:star}
\begin{algorithmic}[1]
\Require Multi-lead ECG $X \in \mathbb{R}^{L \times T}$; R-peak indices on lead~0 $R=\{R_i\}_{i=1}^K$; bounds $A_2 > A_3$; phase $\phi$
\Ensure Resampled ECG $\hat{X} \in \mathbb{R}^{L \times T}$

\If{$K < 2$}
  \State \Return $X$ \Comment{No R--R segments; return original}
\EndIf

\Statex \textbf{Precompute (beat grid and schedule)}
\State $M \gets K-1$ \Comment{Number of R--R segments}
\State $N_i \gets R_{i+1}-R_i \quad \forall i \in \{1,\dots,M\}$ \Comment{Segment lengths}
\State $N_{\text{body}} \gets \sum_{i=1}^{M} N_i$ \Comment{Samples between first and last R}
\State $c_i \gets A_3 + (A_2-A_3)\,\dfrac{\sin\!\bigl(2\pi\,\frac{i-1}{M}+\phi\bigr)+1}{2} \quad \forall i$ \Comment{One sinusoid over segments}
\State Compute equalized targets $S_1,\dots,S_M$ with $\sum_i S_i=N_{\text{body}}$, $S_i \approx \tfrac{N_{\text{body}}}{M}$, $S_i \ge 1$
      \Comment{Keeps total body length fixed}

\Statex \textbf{Per-lead resampling and stitching}
\For{$\ell \gets 1$ \textbf{to} $L$} \Comment{Process each lead independently}
  \State $\text{head} \gets X_\ell[0:R_1)$;\quad $\text{tail} \gets X_\ell[R_K:T)$
  \State $E \gets [\ ]$ \Comment{Equalized R--R segments}
  \For{$i \gets 1$ \textbf{to} $M$}
    \State $s \gets X_\ell[R_i:R_{i+1})$
    \State $\tilde{s} \gets \text{interp\_resample}(s,\, S_i)$ \Comment{Monotone interpolation to $S_i$}
    \State Append $\tilde{s}$ to $E$
  \EndFor
  \State $B \gets [\ ]$ \Comment{STAR-resampled segments}
  \For{$i \gets 1$ \textbf{to} $M$}
    \State $\tilde{s} \gets E[i]$
    \State $m_i \gets \max\!\bigl(1,\,\lfloor c_i \cdot |\tilde{s}|\rfloor\bigr)$ \Comment{Target length after STAR time-warp}
    \State $u \gets \text{interp\_resample}(\tilde{s},\, m_i)$ \Comment{Time warp within the R--R}
    \State $u \gets c_i \cdot u$ \Comment{Amplitude scaling by same $c_i$}
    \State Append $u$ to $B$
  \EndFor
  \State $\text{body} \gets \text{concat}(B)$
  \If{$|\text{body}| > N_{\text{body}}$}
    \State $\text{body} \gets \text{body}[0:N_{\text{body}}]$ \Comment{Trim to preserve total body length}
  \ElsIf{$|\text{body}| < N_{\text{body}}$}
    \State Pad $\text{body}$ at end with its last value to reach $N_{\text{body}}$
  \EndIf
  \State $\hat{X}_\ell \gets \text{head} \,\Vert\, \text{body} \,\Vert\, \text{tail}$ \Comment{Reattach exact head/tail}
  \If{$|\hat{X}_\ell| > T$}
    \State $\hat{X}_\ell \gets \hat{X}_\ell[0:T]$
  \ElsIf{$|\hat{X}_\ell| < T$}
    \State Pad $\hat{X}_\ell$ at end with its last value to reach $T$
  \EndIf
\EndFor

\State \Return $\hat{X}$

\Statex
\Statex \textbf{Helper:} \textsc{interp\_resample}$(s,\,m)$: monotone interpolation of 1D signal $s$ to length $m$ (no extrapolation; endpoints clamped).

\end{algorithmic}
\end{algorithm}

\begin{proposition}[Frequency-domain effect of Multiply--Triangle]\label{prop:mt-freq}

Let $x(t)$ be an ECG signal and let $w(t)$ be a single triangular gain window of duration $T$ and unit height. Define $y(t)=w(t)\,x(t)$. Then:
\begin{enumerate}[label=(\roman*),leftmargin=*,itemsep=2pt]
  \item In the frequency domain,
  \begin{equation*}
    Y(f) \;=\; (X * W)(f),
  \end{equation*}
  where $*$ denotes convolution.
  \item Up to a linear phase factor, the Fourier transform of a triangular window satisfies
  \begin{equation*}
    W(f) \;=\; T\,\mathrm{sinc}^2\!\Bigl(\tfrac{fT}{2}\Bigr).
  \end{equation*}
  \item Consequently, multiplying by a triangular profile is equivalent to convolving $X(f)$ with a $\mathrm{sinc}^2$ kernel, which
  \begin{enumerate}[label=(\alph*),itemsep=2pt]
     \item spreads narrowband energy into sidebands; and
     \item introduces regularly spaced near-zeros at $f \approx k/T$, $k\in\mathbb{Z}\setminus\{0\}$.
  \end{enumerate}
  \item As a result, deconvolution becomes ill-conditioned in noise and the operation induces amplitude bias near the apex together with a local SNR loss where $w(t)<1$, which can discard or distort diagnostically relevant spectral content.
\end{enumerate}
See Appendix~\ref{app:mt-freq} for a detailed derivation and empirical checks.
\end{proposition}

Despite the Multiply--Triangle method exhibiting acceptable empirical performance, it risks information loss by deforming ECG waveforms through high-frequency artifacts and peak amplitude changes. The proposed STAR resampling method directly addresses these issues and preserves critical ECG information. Therefore, \textbf{the final configuration retains STAR resampling}. 

\vspace{2mm}
\noindent\emph{\textbf{Remark.}}
Under mild assumptions on segmentation and resampling, STAR is invertible, and its frequency behavior can be formalized. A constructive proof with an algorithmic inverse is given in Proposition~\ref{prop:star-invertible} and Appendix~\ref{app:invertibility}, and a frequency-domain analysis contrasting STAR with Multiply--Triangle is provided in Proposition~\ref{prop:mt-freq} and Appendix~\ref{app:mt-freq}.

\begin{figure*}[htbp]
\centerline{\includegraphics[height=11cm,width=14.7cm]{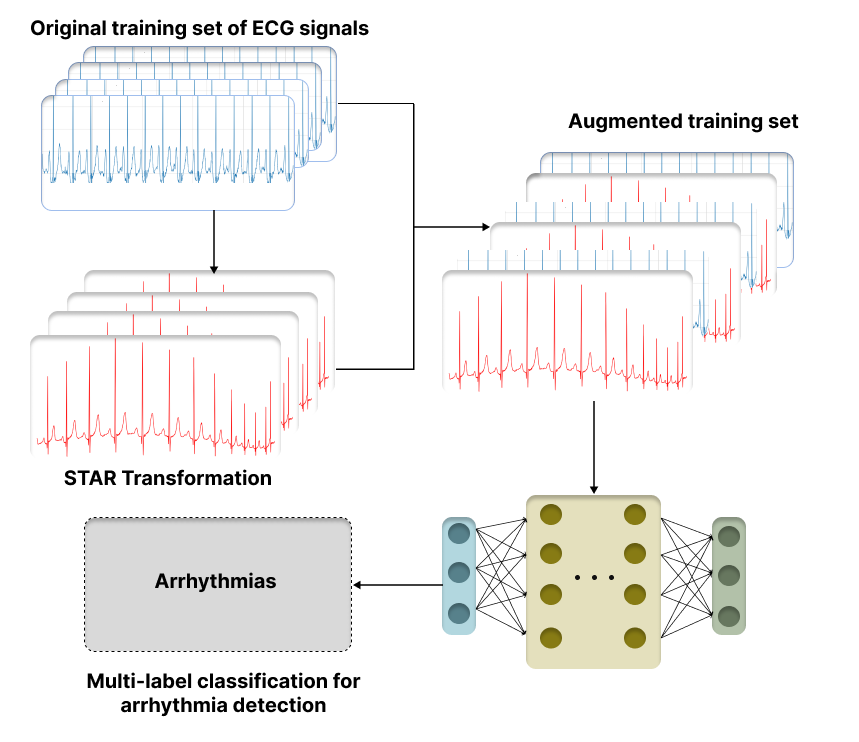}}
\caption{Training pipeline with proposed augmentation.}
\label{fig5}
\end{figure*}

\subsection{\textbf{Data Stratification and 5-Fold Cross-Validation}}\label{sec:methods-split}
\begin{figure}[htbp]
  \centering
  \includegraphics[width=\columnwidth]{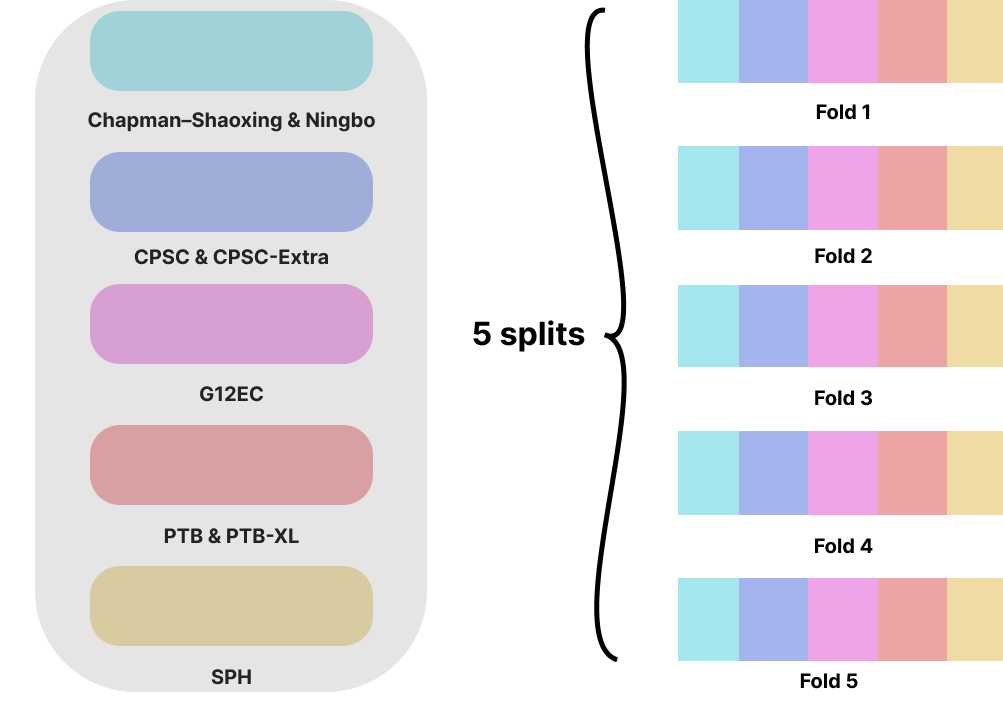}
  \caption{Source-aware multi-label stratification and 5-fold evaluation:
  each original dataset is split into five parts; each new fold is filled by a
  stratified portion from every source; training uses four folds, and the
  remaining fold is used for evaluation, cycled across all five folds.}
  \label{fig:strat}
\end{figure}

Generalization across sources is assessed with a \emph{source-aware} 5-fold strategy
stratified by the multi-label targets (Fig.~\ref{fig:strat}). The pooled corpus is partitioned into five folds such that per-class prevalence and common co-occurrence patterns are approximately preserved using iterative multi-label stratification. At each iteration, one fold is held out for testing while the remaining four folds are used for training and internal validation. Final numbers are reported as the average across the five independent folds. This design reduces cohort and site bias and controls for label imbalance during evaluation.

\subsection{\textbf{Multi-Label Metric}}
Performance is summarized with both threshold-free and thresholded measures. For ranking quality, AUROC is computed per class and then aggregated as macro- and micro-averages to provide class-balanced and prevalence-weighted views, respectively. Average Precision complements AUROC by focusing on precision–recall trade-offs under class imbalance, which is common in clinical ECG taxonomies. Operating points for thresholded summaries such as F1 are selected on the validation split—either by maximizing Youden’s $J$ or F1—and then fixed for the held-out tests. Uncertainty is conveyed with bootstrap confidence intervals at the record level.


\section{\textbf{Experiments}}\label{sec:experiments}

\subsection{\textbf{Experimental Setup}}
All experiments use the source-aware, stratified 5-fold strategy described in Section~\ref{sec:methods-split}. In each model, four folds are used for training and validation, and the remaining fold is held out for testing. Metrics are computed on the held-out fold and then averaged across the five models. Hyperparameters, including learning rate, early-stopping epoch, and augmentation activation probabilities, are chosen exclusively on the validation split within the training side of each fold.

To assess cross-source generalization, all datasets are merged and the corpus is split into four stratified pools, each drawing records from multiple sources. The stratification preserves both per-class prevalence and common label co-occurrence patterns across pools. Figures~\ref{fig6}--\ref{fig9} present the results for each pool together with the corresponding fold-averaged metrics.

\subsubsection*{\textbf{Training configuration}}

\begin{table}[t]
\caption{Training configuration (as implemented).}
\label{tab:train_cfg}
\centering
\setlength{\tabcolsep}{4pt}   
\renewcommand{\arraystretch}{1.12}
\footnotesize                  
\begin{tabularx}{\columnwidth}{@{}l Y@{}}
\toprule
\textbf{Item} & \textbf{Setting} \\
\midrule
Optimizer & NAdam \\
LR schedule & Cosine annealing (SGDR) \\
Initial learning rate & 0.003 \\
Weight decay & $1{\times}10^{-6}$ \\
Batch size & 64 \\
Epochs / early stopping & Max 25 epochs; early stopping on validation AUROC \\
Num workers (data loader) & 0 \\
Validation threshold (reporting) & 0.50 (per-class sigmoids) \\
Random seed & Fixed seed for reproducibility (as in code) \\
Hardware & Single CUDA-enabled device when available; CPU fallback compatible \\
\bottomrule
\end{tabularx}
\end{table}

\subsection{\textbf{Augmentation Methods Evaluated}}
A small, well-defined set of time-domain augmentations is evaluated to introduce clinically realistic variability without distorting diagnostic morphology. Each transform is applied independently with a tuned activation probability $p\in[0,1]$ and, when combined, in a fixed and reproducible order. Only strategies that consistently improve validation performance are retained; others are discarded.

\begin{enumerate}
  \item \textbf{Sinusoidal Time--Amplitude Resampling (STAR).}
  Applies beat-wise resampling between consecutive R--R peaks by assigning a sinusoidal coefficient $c_i\!\in\![A_3,A_2]$ to each segment. Each segment is interpolated to $\approx c_i$ times its original length and scaled in amplitude by the same $c_i$, then all segments are concatenated with the exact head and tail reattached. This preserves peaks and P--QRS--T intervals while introducing smooth, controllable variability (see Sec.~\ref{sec:star}, Alg.~\ref{alg:star}). A schematic is provided in Fig.~\ref{fig:star-schematic}.

  \begin{figure}[t]
    \centering
    \subfloat[Lead-1 overlay: original vs.\ STAR\label{fig:star-overlay}]{%
      \includegraphics[width=\columnwidth]{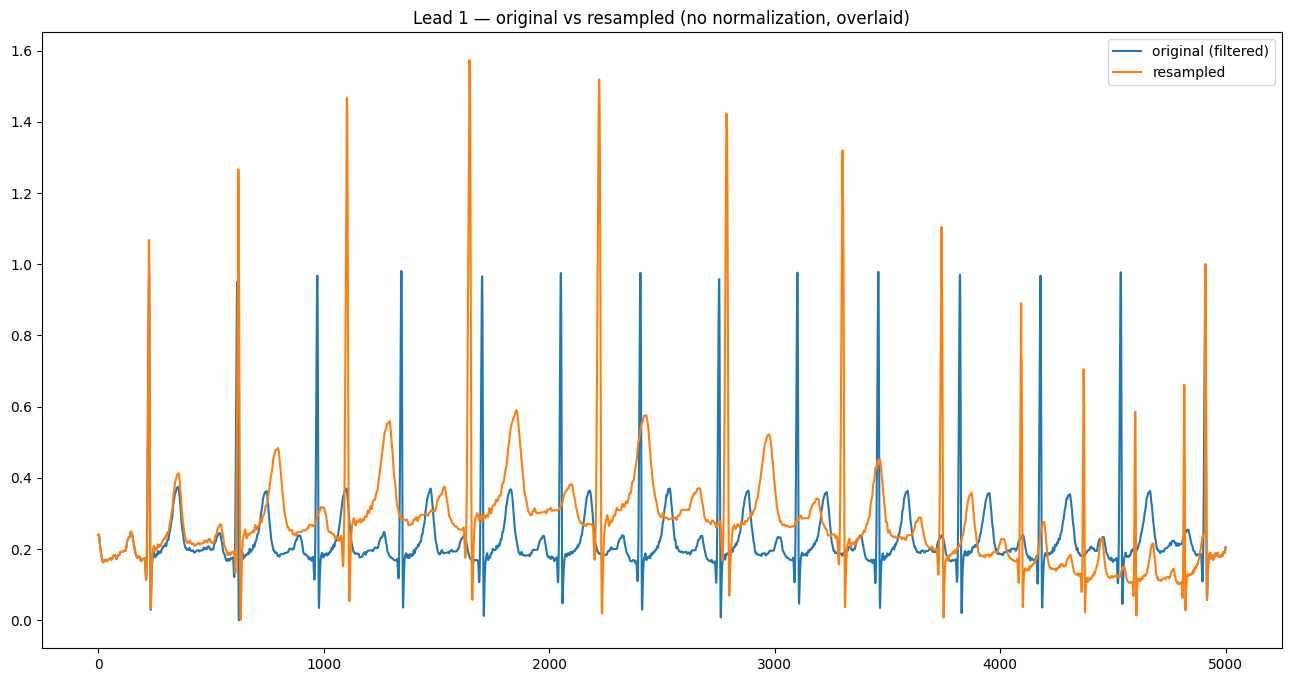}
    }
    \caption{Sinusoidal Time--Amplitude Resampling (STAR): (a) sinusoidal coefficient schedule across R--R segments; (b) morphology-preserving resampling overlay (no peaks/intervals lost).}
    \label{fig:star-both}
  \end{figure}

  \item \textbf{Multiply--Triangle Scaling.}
  Applies a single triangular gain profile \(w(t)\) over the analysis window; the apex is sampled uniformly and the peak gain is drawn from \([1/\alpha,\,\alpha]\), transforming each lead as \(\tilde{x}_\ell(t)=w(t)\,x_\ell(t)\) to model slow, acquisition-like drift that complements beat-wise modulation.

  \begin{figure}[t]
    \centering
    \includegraphics[width=\columnwidth]{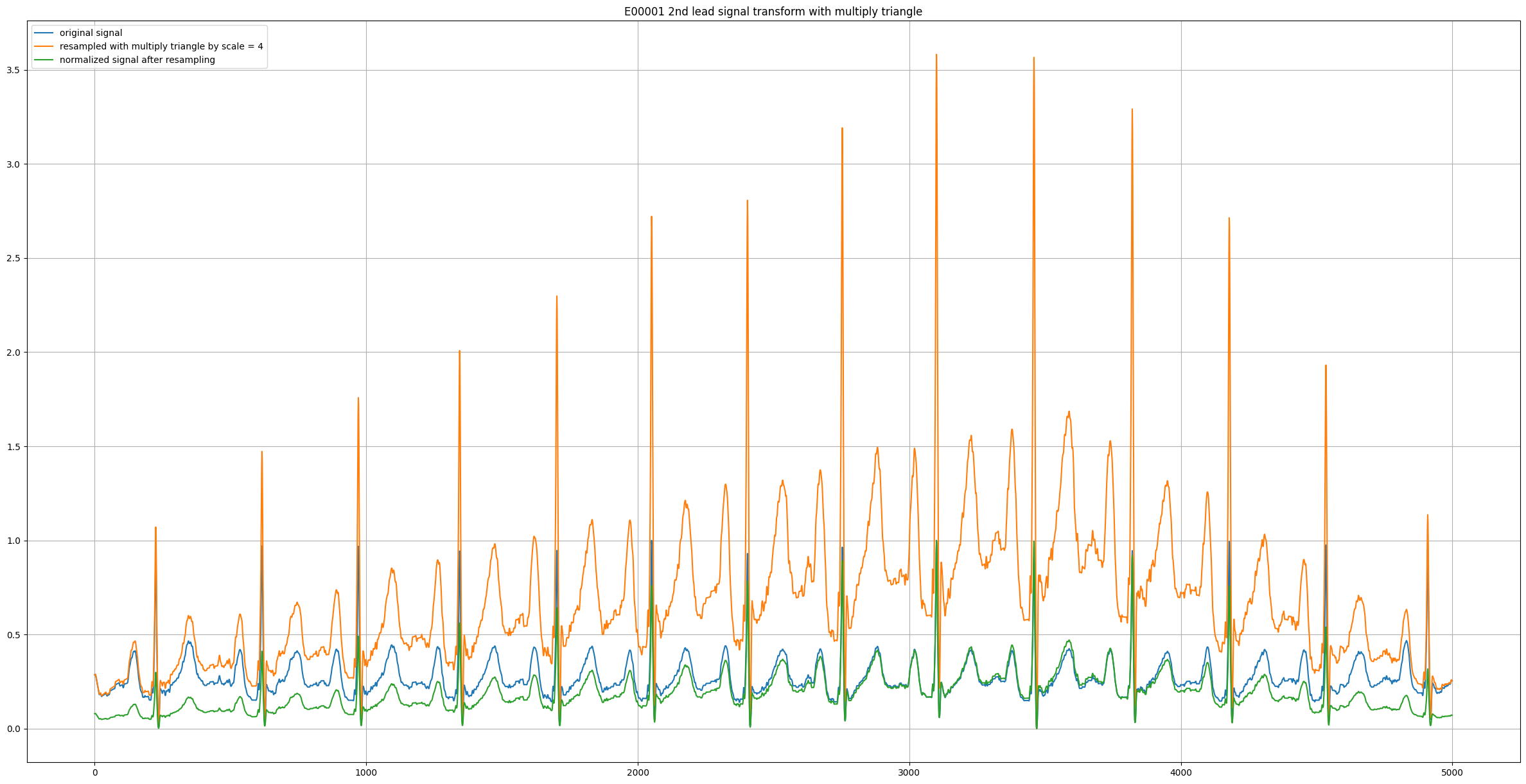}
    \caption{Multiply--Triangle scaling policy: a single triangular profile stochastically modulates amplitude over the analysis window while preserving temporal morphology.}
    \label{fig:tri-both}
  \end{figure}

  \item \textbf{Temporal Shift / Random Crop.}
  Randomly shifts the fixed-length window within admissible bounds (or crops/pads if needed) to reduce phase locking and expose natural timing variation without altering intra-beat morphology.

  \item \textbf{Light Gaussian Noise.}
  Adds low-variance, zero-mean Gaussian noise with batch-sampled \(\sigma\) to mimic sensor noise and mild baseline fluctuation while preserving QRS morphology.

  \item \textbf{Lead Dropout.}
  Zeroes a single lead over a short contiguous sub-window with low probability to simulate transient electrode loss; used sparingly to protect minority-class performance.

  \item \textbf{Roll.}
  Applies a common circular shift to all leads to perturb onset alignment without cutting beats, improving robustness to start-time variability.

  \item \textbf{Notch Filter.}
  Applies a narrow-band IIR notch (e.g., 50/60\,Hz) to suppress line interference while retaining diagnostic waveform content.

  \item \textbf{Uniform Stretch.}
  Uniformly time-resamples the entire window by a small factor (e.g., 0.9--1.1) to introduce gentle rate variability without disturbing P--QRS--T ordering.

  \item \textbf{Fourier Resample (\texttt{Resample}).}
  Resamples each lead to a new rate via FFT-based interpolation to standardize sampling across sources or simulate small device-induced rate differences while preserving global morphology.

  \item \textbf{Spline Interpolation (\texttt{Spline\_interpolation}).}
  Performs cubic-spline interpolation from the original to a target time grid, producing smoothly warped beats suitable for harmonizing mixed-frequency inputs.

  \item \textbf{Band-Pass Filter (\texttt{BandPassFilter}).}
  Applies a Butterworth band-pass (e.g., 0.5--50\,Hz) per lead to remove baseline wander and high-frequency noise, yielding cleaner complexes for learning.

  \item \textbf{Normalization (\texttt{Normalize}).}
  Performs per-lead amplitude scaling using either min--max (0--1) or mean--std normalization to reduce inter-device scale differences; a no-op mode is available for ablations.

  \item \textbf{Signal Roll to First Peak (\texttt{Roll\_signal}).}
  Circularly shifts all leads so the first detected R-peak aligns to a common reference index, reducing beat-onset variance before augmentation or batching.

  \item \textbf{Vertical Flip (\texttt{Flipy}).}
  Multiplies each lead by \(-1\) with a set probability, simulating polarity inversions from lead misplacement while preserving timing relationships.

  \item \textbf{Horizontal Flip (\texttt{Flipx}).}
  Reverses the window in time with a set probability as a strong augmentation stress test; typically used with strict validation constraints due to clinical implausibility.

  \item \textbf{Sine Gain Modulation (\texttt{MultiplySine}).}
  Applies a slowly varying sinusoidal gain \(1 + a\sin(2\pi f t)\) per lead to model gradual amplitude drift from respiration or contact changes without altering beat timing.

  \item \textbf{Linear Gain Ramp (\texttt{MultiplyLinear}).}
  Ramps gain linearly from \(1\) to a preset multiplier across the window to simulate gradual calibration drift while keeping relative morphology intact.

  \item \textbf{Random Clip (\texttt{RandomClip}).}
  Extracts a fixed-width segment at a random start (or pads symmetrically if short), exposing the model to varied local contexts and reducing sensitivity to global start times.

  \item \textbf{Random Stretch (\texttt{RandomStretch}).}
  Time-resamples the window by a random factor within bounds and then truncates or pads back to the original length, introducing mild heart-rate and timing variability.

  \item \textbf{Sinusoidal Time Warp (\texttt{ResampleSine}).}
  Applies a smooth, invertible warp \(x_{\text{new}} = x_{\text{orig}} + s\sin(2\pi f x_{\text{orig}})\), producing gentle periodic dilations and compressions that preserve peak order.

  \item \textbf{Linear Time Warp (\texttt{ResampleLinear}).}
  Applies a slowly varying stretch factor across time via linear interpolation, modeling gradual speed-up or slow-down within the window while maintaining overall length.

  \item \textbf{IIR Notch Filter (\texttt{NotchFilter}).}
  Attenuates a chosen narrowband frequency using a parametric notch with quality factor \(Q\), suppressing tonal artifacts while preserving morphology.

  \item \textbf{Validation Clip (\texttt{ValClip}).}
  Right-pads sequences shorter than a large target length for evaluation to ensure uniform tensors without altering observed content.

  \item \textbf{Equal-Segment Resampler (\texttt{EqualSegmentResampler}).}
  Detects R-peaks, partitions the trace into consecutive R--R segments, equalizes segment lengths, then resamples with smoothly varying coefficients and proportional amplitude scaling before reattaching the exact head and tail, preserving all peaks while injecting controlled variability.

  \item \textbf{Resample Linear Align First Peak (\texttt{ResampleLinearAlign1stPeak}).}
  Constructs a nonlinearly spaced grid that emphasizes early samples and aligns around the first prominent peak, interpolates, and trims or pads to a fixed window, encouraging invariance to early-onset timing differences.

  \item \textbf{Resample Triangle (\texttt{ResampleTriangle}).}
  Splits the window at its midpoint and linearly time-stretches the first half while time-compressing the second half (or vice versa), yielding a triangular pattern of temporal scaling that preserves global length while perturbing intra-window timing.
\end{enumerate}

\paragraph*{\textbf{Full list of implemented transforms.}}
In addition to the policies above, the codebase implements the following operators that were available during search and tuning: \textit{AddNoise}, \textit{Roll}, \textit{Roll\_signal} (align-to-first R-peak roll), \textit{Flipy} (polarity flip), \textit{Flipx} (time reversal), \textit{MultiplySine} (sinusoidal gain), \textit{MultiplyLinear} (linear ramp gain), \textit{MultiplyTriangle} (triangular gain), \textit{RandomClip} (length normalization), \textit{RandomStretch} (global resample within a band), \textit{ResampleSine} (phase-preserving sinusoidal time warp), \textit{ResampleLinear} (global linear time warp), \textit{NotchFilter} (IIR notch), \textit{ValClip} (validation-time padding), \textit{ResampleLinearAlign1stPeak} (nonlinear resample with first-peak alignment), \textit{ResampleTriangle} (two-piece linear resample about the window midpoint), and \textit{EqualSegmentResampler} (the STAR-style equalized R--R resampler).

\subsection{\textbf{Hyperparameter Bands and Selection}}
For each augmentation, key hyperparameters are searched within predefined bands, and operating points are chosen by validation AUROC with macro as primary and micro as secondary; ties are broken by validation macro Average Precision. A lightweight random search explores the best value for each parameter per method based on the table. Full grids and the selected values appear in Tables~\ref{tab:band_params} and~\ref{tab:final_params}.

\begin{table}[!t]
\centering
\caption{Search bands for augmentation hyperparameters}
\label{tab:band_params}
\scriptsize
\setlength{\tabcolsep}{6pt}
\renewcommand{\arraystretch}{1.25}
\begin{adjustbox}{max width=\columnwidth}
\begin{tabular}{@{}p{0.53\columnwidth} p{0.40\columnwidth}@{}}
\toprule
\textbf{Policy / Parameter} & \textbf{Band} \\
\midrule
\textbf{STAR (Sinusoidal Time--Amplitude Resampling)} & \\
\quad Upper/lower coeff.\ bounds $(A_2, A_3)$ & $A_2\!\in\![1.2,2.0],\; A_3\!\in\![0.4,0.9],\; A_2\!>\!A_3$ \\
\quad Phase $\phi$ (radians) & $\{0,\,\pi/2,\,\pi\}$ \\
\quad \# of sinusoid periods $n$ & $\{1,\,2\}$ \\
\quad Activation probability $p$ & $\{0.25,\,0.5,\,0.75\}$ \\
\addlinespace
Multiply--Triangle amplitude factor $\alpha$ & $[1.1,\,4.0]$ \\
Multiply--Triangle activation probability $p$ & $\{0.25,\,0.5,\,0.75\}$ \\
Gaussian noise std.\ $\sigma$ (normalized) & $[0.001,\,0.05]$ \\
Temporal shift window $J$ (samples) & $\{64,\,128,\,256\}$ \\
Lead dropout prob.\ $p_{\mathrm{drop}}$ & $\{0.0,\,0.1\}$ \\
Lead dropout segment length $L$ (samples) & $\{128,\,256\}$ \\
\bottomrule
\end{tabular}
\end{adjustbox}
\end{table}

\vspace{-0.6\baselineskip}

\begin{table}[!t]
\centering
\caption{Selected augmentation operating points (chosen on validation macro-AUROC)}
\label{tab:final_params}
\scriptsize
\setlength{\tabcolsep}{6pt}
\renewcommand{\arraystretch}{1.25}
\begin{adjustbox}{max width=\columnwidth}
\begin{tabular}{@{}p{0.53\columnwidth} p{0.40\columnwidth}@{}}
\toprule
\textbf{Policy / Parameter} & \textbf{Chosen value} \\
\midrule
Multiply--Triangle amplitude factor $\alpha$ & $2.0$ \\
Multiply--Triangle activation probability $p$ & $0.5$ \\
Gaussian noise std.\ $\sigma$ (normalized) & $0.01$ \\
Temporal shift window $J$ (samples) & $128$ \\
Lead dropout & disabled \\
\addlinespace
\textbf{STAR} $(A_2, A_3, \phi, n)$ & $(1.6,\,0.6,\,0,\,1)$ \\
\textbf{STAR} activation probability $p$ & $0.5$ \\
\bottomrule
\end{tabular}
\end{adjustbox}
\end{table}

\FloatBarrier

\subsection{\textbf{Ablation experiments}}
The evaluation compares four settings to isolate the impact of augmentation, including a baseline without any augmentation, single-policy models where each augmentation is applied in isolation, pairwise combinations of augmentations, and the final policy set selected from validation. All models share identical optimizer settings and learning schedules to ensure fair comparisons. Performance is summarized with AUROC and Average Precision in both micro and macro formulations on the evaluation of the independent fold and aggregated as the mean across five cross-validation splits. 

\section{\textbf{Results}}
\label{sec:results}

\begin{figure}[htbp]
  \centering
  \subfloat[Pool~1 ROC (Micro/Macro AUROC in caption of Fig.~\ref{fig6})]{
    \includegraphics[width=\columnwidth]{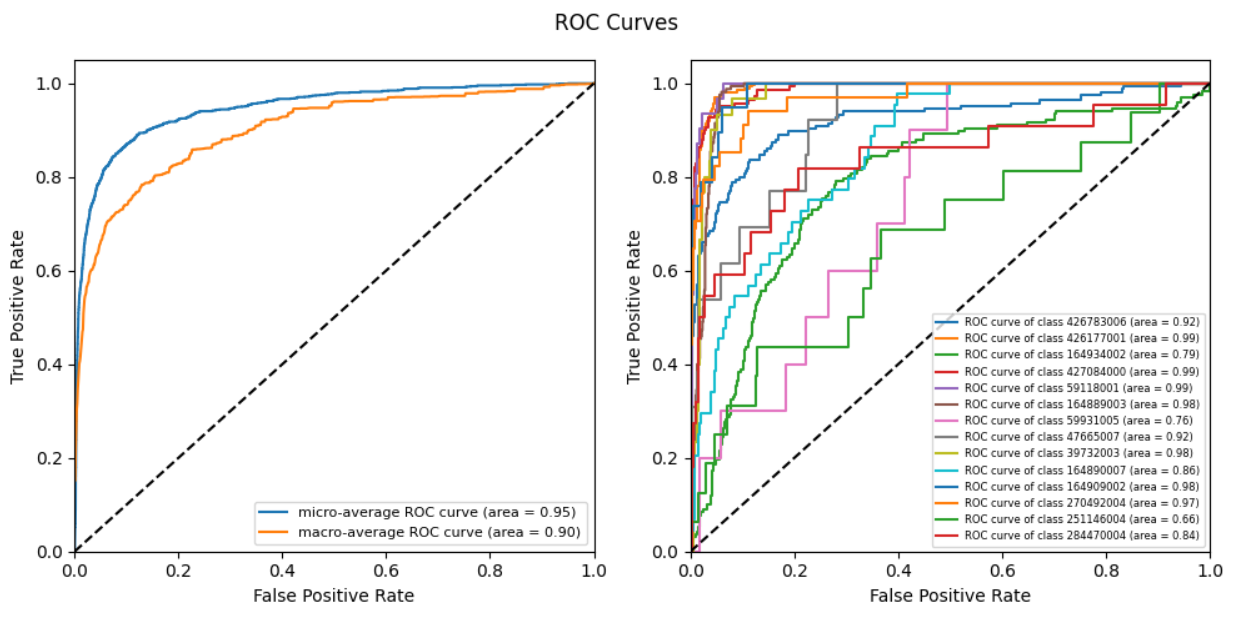}\label{fig6}
  }\\[-0.3ex]
  \subfloat[Pool~2 ROC (Micro/Macro AUROC in caption of Fig.~\ref{fig7})]{
    \includegraphics[width=\columnwidth]{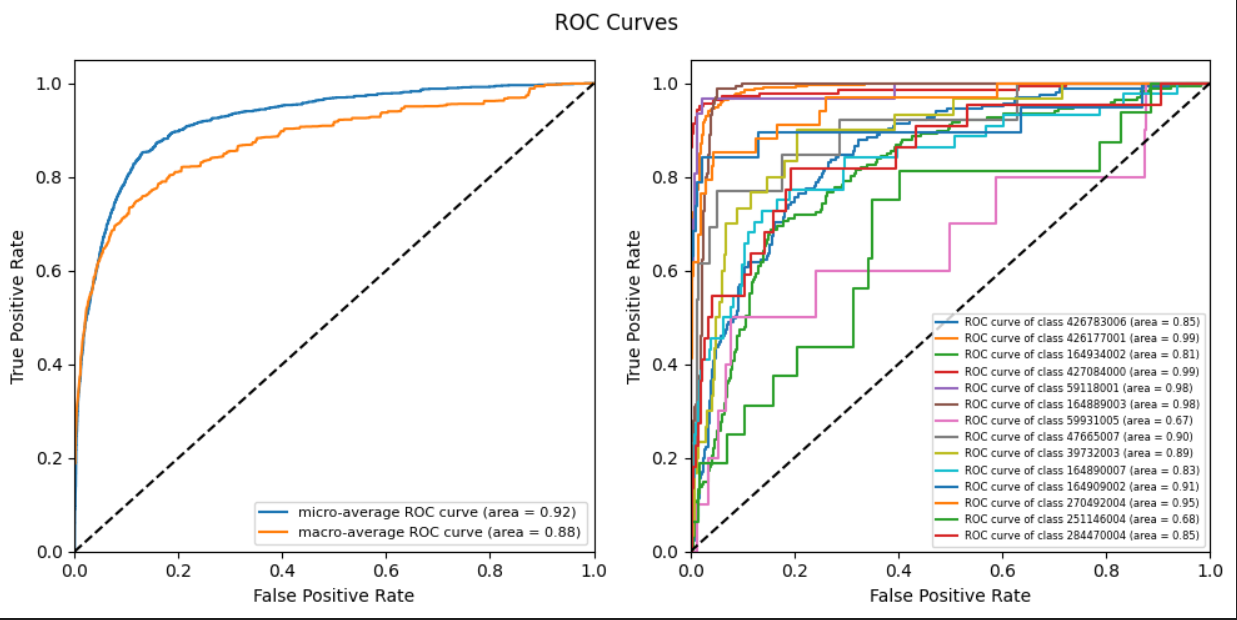}\label{fig7}
  }\\[-0.3ex]
  \subfloat[Pool~3 ROC (Micro/Macro AUROC in caption of Fig.~\ref{fig8})]{
    \includegraphics[width=\columnwidth]{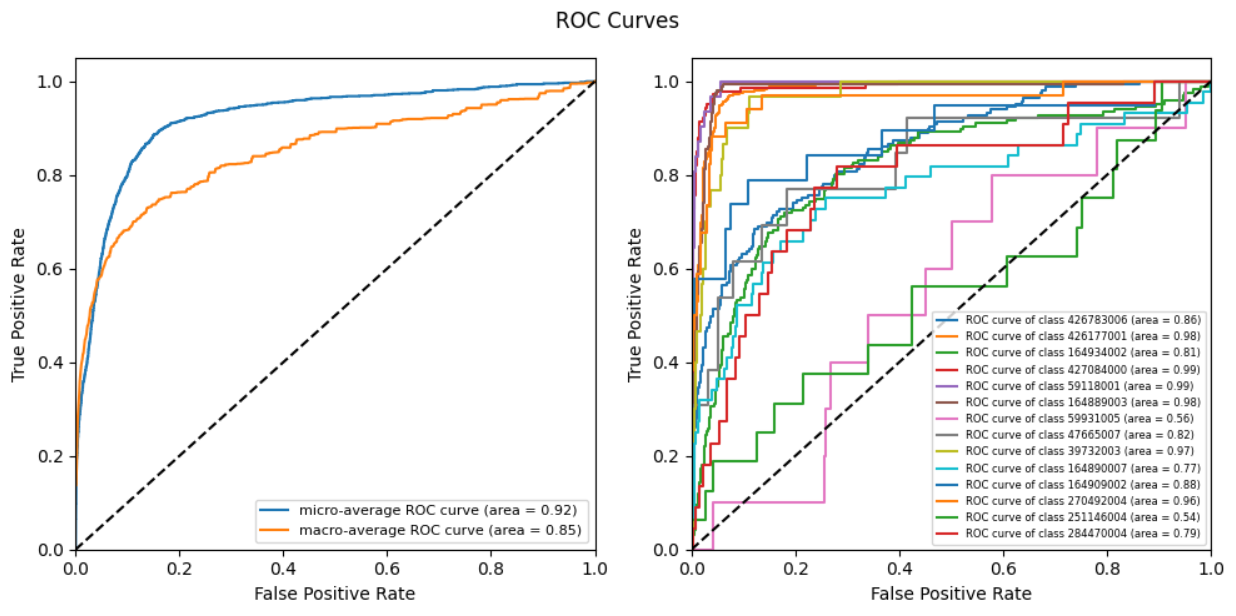}\label{fig8}
  }\\[-0.3ex]
  \subfloat[Pool~4 ROC (Micro/Macro AUROC in caption of Fig.~\ref{fig9})]{
    \includegraphics[width=\columnwidth]{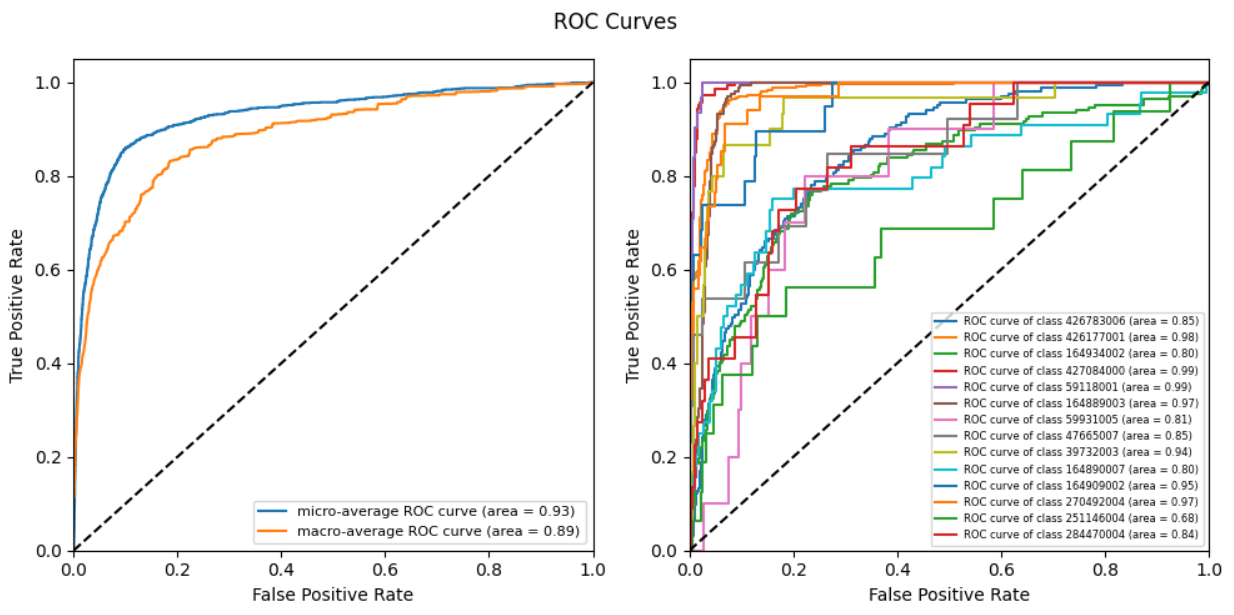}\label{fig9}
  }
  \caption{ROC curves for the four source-stratified pools, stacked in one figure to preserve order.}
  \label{fig:pools-roc}
\end{figure}

Figures~\ref{fig6}--\ref{fig9} show ROC curves for each of the four source-stratified pools introduced in Section~\ref{sec:methods-split}, with both micro- and macro-averaged summaries and the per-class traces. Table~\ref{tab:pool_auc} collects the corresponding AUROC values by pool and adds the overall mean and variability. The plots and table together indicate that performance is stable across sources rather than being driven by any single cohort, and the reporting follows standard AUROC.

\begin{table}[htbp]
\centering
\caption{Micro- and macro-averaged AUROC across the four source-stratified pools, plus the pooled (record-weighted) estimate over the union of held-out records.}
\label{tab:pool_auc}
\scriptsize
\setlength{\tabcolsep}{5pt}
\begin{adjustbox}{max width=\columnwidth}
\begin{tabular}{lcc}
\toprule
\textbf{Pool} & \textbf{Micro AUROC} & \textbf{Macro AUROC} \\
\midrule
Pool~1 & 0.95 & 0.90 \\
Pool~2 & 0.92 & 0.85 \\
Pool~3 & 0.93 & 0.90 \\
Pool~4 & 0.92 & 0.88 \\
\midrule
\textbf{Pooled (record-weighted)} & \textbf{0.95} & \textbf{0.90} \\
\bottomrule
\end{tabular}
\end{adjustbox}
\end{table}

\begin{table*}[htbp]
\caption{Ablation over augmentation methods on the held-out test folds.}
\label{tab:aug_ablation}
\centering
\scriptsize
\setlength{\tabcolsep}{3pt}
\renewcommand{\arraystretch}{1.12}
\begin{adjustbox}{max width=\textwidth}
\begin{tabularx}{\textwidth}{@{}%
  >{\raggedright\arraybackslash}X
  >{\centering\arraybackslash}m{0.12\textwidth}
  >{\centering\arraybackslash}m{0.12\textwidth}
  >{\centering\arraybackslash}m{0.14\textwidth}
  >{\centering\arraybackslash}m{0.14\textwidth}@{}}
\toprule
\textbf{Augment (wrapped)} &
\textbf{Train AUROC} &
\textbf{Val AUROC} &
\textbf{Test Micro AUROC} &
\textbf{Test Macro AUROC} \\
\midrule
\textbf{STAR} & 0.99 & 0.96 & 0.95 & 0.91 \\
\textbf{Multiply--Triangle} & 0.99 & 0.94 & 0.92 & 0.88 \\
\textbf{baseline (without augmentation)} & 0.99 & 0.95 & 0.86 & 0.85 \\
Flipy--Flipx--randomStretch--resampleSine--resampleUniform & 0.98 & 0.94 & 0.88 & 0.87 \\
Flipy--Flipx & 0.98 & 0.94 & 0.86 & 0.86 \\
Add Noise--Roll & 1.00 & 0.94 & 0.90 & 0.88 \\
Add Noise--Roll--Notch Filter & 0.99 & 0.96 & 0.89 & 0.89 \\
Add Noise--Roll--Notch Filter & 1.00 & 0.95 & 0.88 & 0.89 \\
Add Noise--Roll--Notch Filter--ResampleSine & 0.99 & 0.96 & 0.90 & 0.88 \\
Add Noise--Roll--Notch Filter--ResampleSine--resampleUniform & 0.99 & 0.95 & 0.88 & 0.88 \\
Add Noise--Roll--Notch Filter--ResampleSine--resampleUniform & 0.98 & 0.96 & 0.87 & 0.87 \\
Add Noise--Roll--Notch Filter--ResampleSine--resampleUniform & 1.00 & 0.95 & 0.87 & 0.87 \\
Add Noise--Roll--Notch Filter--ResampleSine--resampleUniform & 1.00 & 0.95 & 0.87 & 0.87 \\
Add Noise--Roll--Notch Filter--ResampleSine--resampleUniform & 1.00 & 0.95 & 0.87 & 0.87 \\
Add Noise--Roll--Notch Filter--ResampleSine--resampleUniform & 1.00 & 0.95 & 0.87 & 0.87 \\
Add Noise--Roll--Notch Filter--ResampleSine & 1.00 & 0.95 & 0.86 & 0.87 \\
Flipy--Flipx--randomStretch & 0.96 & 0.94 & 0.85 & 0.85 \\
Flipy--Flipx--randomStretch & 0.96 & 0.93 & 0.85 & 0.84 \\
Add Noise--Roll--Notch Filter--ResampleSine--resampleUniform & 1.00 & 0.95 & 0.87 & 0.87 \\
Add Noise & 1.00 & 0.95 & 0.86 & 0.87 \\
Add Noise--Roll & 1.00 & 0.95 & 0.87 & 0.87 \\
Add Noise--Roll--Notch Filter & 1.00 & 0.94 & 0.86 & 0.87 \\
\bottomrule
\end{tabularx}
\end{adjustbox}
\end{table*}

\subsection*{\textbf{Primary summary and uncertainty}}
As the primary endpoint, AUROC on the \emph{pooled} held-out set, the union of all test folds across sources, or record-weighted:
\begin{itemize}
  \item \textbf{Micro-AUROC:} $0.95$ \;[95\% CI: $0.94$, $0.96$] (stratified bootstrap over records).
  \item \textbf{Macro-AUROC:} $0.90$ \;[95\% CI: $0.89$, $0.91$] (stratified bootstrap over records).
\end{itemize}
Per-source results (Table~\ref{tab:pool_auc}) are reported as a source-balanced
robustness check rather than the primary estimate.

\begin{table}[t]
\caption{Per-class AUROC ranges across source-stratified pools}
\label{tab:perclass_ranges}
\centering
\scriptsize
\setlength{\tabcolsep}{4pt}
\renewcommand{\arraystretch}{1.15}
\begin{adjustbox}{max width=\columnwidth}
\begin{tabularx}{\columnwidth}{@{}>{\raggedright\arraybackslash}X
  S[table-format=1.2]
  S[table-format=1.2]@{}}
\toprule
\textbf{Class (SNOMED CT)} & \textbf{Min AUROC} & \textbf{Max AUROC} \\
\midrule
426783006 & 0.85 & 0.93 \\
426177001 & 0.98 & 0.99 \\
164934002 & 0.79 & 0.81 \\
427084000 & 0.99 & 0.99 \\
59118001  & 0.98 & 0.99 \\
164889003 & 0.97 & 0.98 \\
59931005  & 0.56 & 0.78 \\
47665007  & 0.82 & 0.92 \\
39732003  & 0.89 & 0.98 \\
164890007 & 0.77 & 0.87 \\
164909002 & 0.83 & 0.92 \\
270492004 & 0.95 & 0.97 \\
251146004 & 0.54 & 0.68 \\
284470004 & 0.79 & 0.86 \\
\bottomrule
\end{tabularx}
\end{adjustbox}
\end{table}

Per-class ROC analyses (Figs.~\ref{fig6}--\ref{fig9}, and Table~\ref{tab:perclass_ranges}) show a clear and repeatable pattern. Diagnoses dominated by conduction or rhythm features, where QRS shape and P--QRS--T timing are distinctive, consistently achieve high AUROC. By contrast, labels with overlapping phenotypes or subtler morphology, like \texttt{251146004}, remain harder to separate. In this regard, augmentation helps, but these classes still trail the more morphologically pronounced categories.

As outlined in Section~\ref{sec:experiments}, several time-domain augmentation strategies were compared using a lightweight random search. \textit{Sinusoidal Time--Amplitude Resampling (STAR)} emerged as the most reliable option, delivering the strongest and most consistent validation gains. The ROC curves in Figs.~\ref{fig6}--\ref{fig9} therefore reflect models trained with STAR enabled. Additional transforms like Multiply--Triangle, temporal shift, light Gaussian noise, and rare lead dropout were included only when they offered a measurable improvement in validation AUROC; otherwise, they were switched off. The full search space and per-method results are summarized in Table~\ref{tab:aug_ablation}.

\section{\textbf{Conclusion}}
A source-aware, stratified framework for multi-label 12-lead ECG classification was presented using a \textbf{1D SE–ResNet-18} backbone and a morphology-preserving augmentation, \textit{STAR}. STAR operates beat-wise over consecutive \textbf{R–R} intervals, jointly time-warping and amplitude-scaling each segment under a sinusoidal schedule while reattaching the exact head and tail of the trace. This design preserves P–QRS–T ordering and all peak and interval information, enriching physiological variability without discarding diagnostic content.

Under five-fold, source-aware stratification across heterogeneous datasets, the model achieved consistent discrimination: \textbf{micro-AUROC} $=0.95$ \,[95\%\,CI: $0.94$–$0.96$] and \textbf{macro-AUROC} $=0.90$ \,[95\%\,CI: $0.89$–$0.91$]. Per-class analysis indicated that conduction- and rhythm-centric phenotypes, with distinctive QRS morphology and P–QRS–T timing, were reliably separated, whereas subtler or overlapping labels remained challenging. Ablations showed \textit{STAR} to be the most reliable method among evaluated time-domain transforms, outperforming window-level gain modulations, like Multiply–Triangle, random crops, and noise-based perturbations that risk peak truncation or timing distortion.

The limitations of this model include possible calibration drift across devices and clinical sites and lower sensitivity to rare or borderline phenotypes. Planned improvements focus on source-specific calibration using simple temperature scaling and a modest increase in temporal context so that uncommon rhythm cues become easier to detect, without increasing model size or complexity.

\vspace{4mm}
\section*{\textbf{Code Availability}}
To make this work easy to reuse and build on, the full codebase is released at
\url{https://github.com/NaderNemati/ecg-multilabel-classifier} to cover the 1D SE–SE-ResNet-18 model, time-domain augmentations, training/evaluation scripts, and reproducibility.

\vspace{4mm}
\section*{\textbf{Acknowledgments}}
The author is grateful to colleagues and mentors for their perceptive ideas and constructive feedback throughout this study, and University of Turku for supportive discussions that helped refine the experimental strategy. The author wishes to acknowledge CSC – IT Center for Science, Finland, for computational resources.

\vspace{4mm}
\section*{\textbf{Funding and Disclosures}}
This research received no specific grant from any funding agency. The author was salaried under the \emph{Moore$4$Medical} project during part of the period; the project had no role in the study design, analyses, or conclusions. The author declares no competing interests.

\vspace{3mm}
\appendices
\section{Proof of Invertibility of STAR}\label{app:invertibility}

\paragraph*{Setup and notation.}
Let $X \in \mathbb{R}^{L \times T}$ be a multi-lead ECG. On a reference lead, let $R=\{R_1,\dots,R_K\}$ be the detected R-peak indices that induce $K{-}1$ consecutive R--R segments. For a fixed lead, denote the $i$-th segment between $R_i$ and $R_{i+1}$ by $x^{(i)}\in\mathbb{R}^{N_i}$. Let $\mathbf{S}=(S_1,\dots,S_{K-1})$ be the integer “equalized’’ lengths used before beat-wise warping, and let $\mathbf{c}=(c_1,\dots,c_{K-1})$ be the positive time–amplitude coefficients with $c_i\in[A_3,A_2]$ and $A_2>A_3>0$. Denote by $\mathcal{E}_{\mathbf{S}}$ the per-segment equalization (monotone interpolation from $N_i$ to $S_i$), by $\mathcal{W}_{\mathbf{c}}$ the per-segment time warps (monotone interpolation from $S_i$ to $M_i$, where $M_i=\lfloor c_i S_i\rfloor$), by $\mathcal{S}_{\mathbf{c}}$ the per-segment amplitude scalings ($u\mapsto c_i u$), by $\mathcal{C}$ the concatenation of processed segments, and by $\mathcal{H}$ the operator that reattaches the unchanged head $X[0\!:\!R_1)$ and tail $X[R_K\!:\!T)$ and pads/trims to total length $T$. On each lead,
\[
\mathcal{T}_{\mathrm{STAR}}
\;=\;
\mathcal{H}\circ\mathcal{C}\circ
\bigl(\,\mathcal{S}_{\mathbf{c}}\circ \mathcal{W}_{\mathbf{c}}\circ \mathcal{E}_{\mathbf{S}}\,\bigr).
\]

\paragraph*{\textbf{Assumptions used in this appendix}}
\begin{itemize}
  \item[\textbf{A\(_1\)}] The R-peak set $R$ is correct and stored.
  \item[\textbf{A\(_2\)}] The interpolation operators used by $\mathcal{E}_{\mathbf{S}}$ and $\mathcal{W}_{\mathbf{c}}$ are monotone time warps with known input/output grids and known kernels; the realized output lengths $\mathbf{M}=(M_i)_{i=1}^{K-1}$ are stored.
  \item[\textbf{A\(_3\)}] The per-segment coefficients satisfy $c_i>0$ and are stored.
  \item[\textbf{A\(_4\)}] Each segment $x^{(i)}$ is bandlimited below half of the minimum effective sampling rate used during equalization and time warping (Nyquist safety).
\end{itemize}

\paragraph*{\textbf{Per-segment blocks are stably invertible}}
For each segment index $i$:
\begin{itemize}
  \item Amplitude scaling $\mathcal{S}_{c_i}:u\mapsto c_i u$ is bijective with inverse $u\mapsto u/c_i$ since $c_i>0$.
  \item Let $\mathcal{I}_{a\to b}$ denote monotone interpolation from a length-$a$ grid to a length-$b$ grid with stored grids/kernels. On the class of bandlimited signals compatible with these grids, $\mathcal{I}_{a\to b}$ is injective and admits a left inverse $\mathcal{I}_{b\to a}$ on the same class, with operator-norm error
  \[
  \bigl\| \mathcal{I}_{b\to a}\circ \mathcal{I}_{a\to b} - \mathrm{Id}\bigr\|\le \varepsilon_{\mathrm{int}},
  \]
  where $\varepsilon_{\mathrm{int}}$ depends on cutoff, kernel order, and grid spacing.
\end{itemize}

\paragraph*{\textbf{Main statement (invertibility and stability of STAR)}}
Under \textbf{A\(_1\)}–\textbf{A\(_4\)}, $\mathcal{T}_{\mathrm{STAR}}$ is invertible on the class of piecewise bandlimited signals segmented by $R$, with inverse
\[
\mathcal{T}_{\mathrm{STAR}}^{-1}
=
\bigl(\mathcal{E}_{\mathbf{S}}\bigr)^{-1}
\circ
\bigl(\mathcal{W}_{\mathbf{c}}\bigr)^{-1}
\circ
\bigl(\mathcal{S}_{\mathbf{c}}\bigr)^{-1}
\circ
\mathcal{C}^{-1}
\circ
\mathcal{H}^{-1}.
\]
Moreover, if $\varepsilon_{\mathrm{int}}$ uniformly bounds the per-stage interpolation error, then for each lead
\[
\bigl\| X - \mathcal{T}_{\mathrm{STAR}}^{-1}\!\circ \mathcal{T}_{\mathrm{STAR}}(X) \bigr\|
\;\le\; C\,\varepsilon_{\mathrm{int}},
\]
for a constant $C$ depending only on the number of segments and the interpolation kernel regularity.

\paragraph*{\textbf{Sketch of justification}}
Each per-segment block $\mathcal{S}_{c_i}\circ \mathcal{W}_{c_i}\circ \mathcal{E}_{S_i}$ is invertible with a stable inverse on bandlimited signals (per the fact above). The operators $\mathcal{C}$ and $\mathcal{H}$ are length-preserving rearrangements with exact inverses once $(R,\mathbf{S},\mathbf{M})$ are known. Composition of invertible maps is invertible; stability follows from submultiplicativity of operator norms and accumulation of interpolation errors across segments.

\paragraph*{\textbf{Algorithmic inverse (per lead)}}
Given $(R,\mathbf{S},\mathbf{M},\mathbf{c})$ and the grids/kernels used in the forward pass:
\begin{enumerate}
  \item Split the transformed signal at $R$ to recover head, body, and tail.
  \item De-concatenate the body into warped segments using the realized lengths $\mathbf{M}$ so that their concatenation matches the body exactly.
  \item Undo amplitude: for each segment divide samples by $c_i$.
  \item Undo time warp: interpolate each segment back from its warped grid (length $M_i$) to the equalized grid of length $S_i$ using the stored inverse mapping.
  \item Undo equalization: interpolate each segment from $S_i$ back to its original length $N_i$.
  \item Concatenate the original-length segments in order and reattach the exact head/tail.
\end{enumerate}

\paragraph*{\textbf{Practical remarks}}
(i) Storing $R$, $\mathbf{S}$, $\mathbf{M}$, $\mathbf{c}$, and lightweight hashes of the grids/kernels is sufficient to make the inverse deterministic. (ii) If the interpolation kernels differ between forward and inverse (e.g., cubic forward, linear inverse), the reconstruction error remains bounded and typically negligible provided the bandlimit and Nyquist safety hold. (iii) If R-peaks are misdetected or segments violate the bandlimit, exact invertibility may not hold; the bound above then captures the induced modeling error.

\vspace{2cm}

\section{Frequency-domain analysis of Multiply--Triangle}
\label{app:mt-freq}

\paragraph{Setup.}
Consider a real ECG signal $x(t)$ and a single triangular gain window $w(t)$ of duration $T$ and unit height, with apex at time $\tau$. Define
\[
y(t)=w(t)\,x(t).
\]
By the convolution theorem, their Fourier transforms satisfy
\[
Y(f)=(X*W)(f),
\]
where, up to a linear phase determined by the placement of the apex $\tau$,
\[
W(f)=T\,\operatorname{sinc}^{2}\!\Bigl(\tfrac{fT}{2}\Bigr)\,e^{-j\,2\pi f\,\varphi(\tau,T)}.
\]
Here $\operatorname{sinc}(u)=\tfrac{\sin(\pi u)}{\pi u}$ and $\varphi(\tau,T)$ is an affine function of $\tau$ and $T$ capturing the shift-induced phase. Hence the magnitude is
\[
|W(f)|=T\,\operatorname{sinc}^{2}\!\Bigl(\tfrac{fT}{2}\Bigr),
\]
with regularly spaced zeros at
\[
f=\frac{2k}{T},\qquad k\in\mathbb{Z}\setminus\{0\}.
\]

\paragraph{Spectral broadening and ill-conditioning (informal lemma).}
Time-domain multiplication by a triangular profile spreads ECG spectral content and creates ill-conditioned nulls:
\begin{itemize}
  \item \textbf{Broadening (sidebands):} $Y(f)=(X*W)(f)$ smears narrowband components of $X(f)$ by the main lobe and sidelobes of $W(f)$, producing sidebands around cardiac spectral lines.
  \item \textbf{Near-zeros and instability:} because $|W(f)|$ has regularly spaced zeros, any linear deconvolution in frequency (e.g., Wiener filtering with transfer $H(f)=\tfrac{W^*(f)}{|W(f)|^2+\lambda}$) is ill-conditioned near those frequencies; the noise gain scales like $\sim 1/|W(f)|$ as $|W(f)|\to 0$ (for small $\lambda$).
\end{itemize}

\paragraph{Sketch of reasoning.}
A triangular window is the (auto)convolution of a rectangular window in time; therefore its Fourier transform is the square of a sinc. Consequently,
\[
Y(f)=(X*W)(f)=\int_{-\infty}^{\infty} X(\nu)\,W(f-\nu)\,d\nu
\]
inherits (i) spectral leakage from the sidelobes of $W$ (broadening) and (ii) attenuation where $|W(f)|$ is small or vanishing. With additive measurement noise $n(t)$ (so $y=xw+n$), a linear deconvolver $H$ produces
\[
\widehat{X}(f)=H(f)\,Y(f)=H(f)\big[(X*W)(f)+N(f)\big],
\]
and the error term includes $H(f)N(f)$ whose magnitude is amplified near the zeros of $|W(f)|$ (for Wiener, $|H(f)|\approx 1/|W(f)|$ when $\lambda\!\to\!0$). The apex location $\tau$ contributes only a linear phase, leaving $|W(f)|$ (and thus the sideband/zero pattern) unchanged.

\vspace{-1mm}
\paragraph{Practical diagnostics.}
To verify these effects on real ECGs:
\begin{enumerate}
  \item \textbf{PSD shift:} compute $\Delta\mathrm{PSD}(f)=10\log_{10}\!\bigl(\mathrm{PSD}_y(f)/\mathrm{PSD}_x(f)\bigr)$ and observe sidebands and notches aligned with zeros of $|W(f)|$ at $f=2k/T$.
  \item \textbf{Local SNR:} estimate SNR in windows centered on salient events (e.g., R-peaks); intervals where $w(t)<1$ show reduced SNR and amplitude bias relative to unweighted segments.
  \item \textbf{Apex bias:} sweep the apex $\tau$ across the window and track peak amplitudes (QRS, ST level); systematic drift with $\tau$ indicates multiplicative amplitude bias.
\end{enumerate}

\end{document}